\documentclass[screen, acmsmall]{acmart}

\def\BibTeX{{\rm B\kern-.05em{\sc i\kern-.025em b}\kern-.08em
    T\kern-.1667em\lower.7ex\hbox{E}\kern-.125emX}}

\usepackage{colortbl}
\usepackage{graphicx}
\usepackage{listings}
\usepackage[shortlabels]{enumitem}
\usepackage[ruled,vlined,linesnumbered]{algorithm2e}
\usepackage{amsthm}
\usepackage{xspace}
\usepackage{soul}
\usepackage{multirow}
\usepackage{makecell}
\usepackage{pgfplots}
\usepackage{threeparttable}
\usepackage{wasysym}
\usepackage{xparse}
\usepackage{microtype}
\usepackage{textcomp}
\usepackage{xcolor}
\usepackage{cleveref}
\usepackage{subfigure}
\usepackage{marvosym}
\usepackage{booktabs} 
\usepackage{tabularx}

\newcommand{\toolName}{\textsc{Legion}\xspace}
\newcommand{\toolNameA}{\textsc{Legion}$^\alpha$\xspace}
\newcommand{\toolNameAshort}{\textsc{Lg}$^\alpha$\xspace}
\newcommand{\toolNameB}{\textsc{Legion}$^\beta$\xspace}
\newcommand{\toolNameBshort}{\textsc{Lg}$^\beta$\xspace}
\newcommand{\toolNameC}{\textsc{RS}\xspace}
\newcommand{\toolNameD}{\textsc{Cov}\xspace}
\newcommand{\toolNameE}{\textsc{EA}\xspace}
\newcommand{\toolNameF}{\textsc{WSY}\xspace}

\newcommand{\increase}{{\color{red}\rotatebox{90}{\MVRightarrow}}}

\newboolean{showcomments}
\setboolean{showcomments}{true}
\ifthenelse{\boolean{showcomments}}
 { \newcommand{\mynote}[2]{
      \fbox{\bfseries\sffamily\scriptsize#1}
        {\small$\blacktriangleright$\textsf{\emph{#2}}$\blacktriangleleft$}}}
        { \newcommand{\mynote}[2]{}}

\newcommand{\tuple}[1]{\langle #1 \rangle}

\definecolor{dark-gray}{gray}{0.45}

\newcommand{\rowsh}{\rowcolor{gray!15}}

\SetKwProg{Fn}{Function}{}{end}
\SetKwFunction{FRecurs}{FnRecursive}
\SetKwProg{Try}{Try}{}{}
\SetKw{Catch}{CatchAndIngore}
\SetKw{Throw}{throw exception}
\SetKw{Continue}{continue}
\SetKw{Break}{break}

\definecolor{darkgray}{rgb}{0.4, 0.4, 0.4}

\SetCommentSty{mycommfont}

\usepackage{tcolorbox}
\tcbset{colback=blue!5!white,colframe=blue!75!black,fonttitle=fseries,size=fbox}

\usepackage{tikz}
\usetikzlibrary{arrows,fit,positioning,shapes.misc}

\tikzset{%
  highlight/.style={rectangle,rounded corners,fill=red!15,draw,fill opacity=0.5,thick,inner sep=0pt}
}

\renewcommand\footnotetextcopyrightpermission[1]{}
\setcopyright{none}

\usepackage{array}

\newcolumntype{L}[1]{>{\raggedright\let\newline\\\arraybackslash\hspace{0pt}}m{#1}}
\newcolumntype{C}[1]{>{\centering\let\newline\\\arraybackslash\hspace{0pt}}m{#1}}
\newcolumntype{R}[1]{>{\raggedleft\let\newline\\\arraybackslash\hspace{0pt}}m{#1}}

\usepackage{pifont}

\newcommand{\greyboxb}[2]{
\vspace{0.05cm}
    \begin{tcolorbox}[
        left=2pt, right=2pt, top=2pt, bottom=2pt,
        boxrule=0.2mm,
        leftrule=2mm,
        arc=0mm,
        colframe=black!40!white,
        colback=black!5!white,
        colbacktitle=black!50!white
    ]
    \textbf{#1}{#2}
    \end{tcolorbox}
}

\begin{document}
	
\title{Ensemble Fuzzing with Dynamic Resource Scheduling and Multidimensional Seed Evaluation}

\newcommand{\zju}{The State Key Laboratory of Blockchain and Data Security, Zhejiang University} 

\author{Yukai Zhao}
\authornote{The two lead authors contributed equally to this work}
\orcid{0009-0000-5366-6349} 
\affiliation{    
  % \institution{\zju}
  \institution{School of Software Technology, Zhejiang University}
  \city{Ningbo}
  \country{China}
}
\affiliation{
  \institution{\zju}
  \city{Hangzhou}
  \country{China}
}
\email{yukaizhao2000@zju.edu.cn}

\author{Shaohua Wang}
\authornotemark[1]
\orcid{0000-0001-5777-7759} 
\affiliation{    
  \institution{Central University of Finance and Economics}
  \city{Beijing}
  \country{China}
}
\email{davidshwang@ieee.org}

\author{Jue Wang}
\orcid{0000-0001-9866-2624} 
\affiliation{    
  \institution{Nanjing University}
  \city{Nanjing}
  \country{China}
}
\email{juewang591@gmail.com}

\author{Xing Hu}
\authornote{Corresponding Author}
\orcid{0000-0003-0093-3292} 
\affiliation{
  \institution{\zju}
  \city{Hangzhou}
  \country{China}
}
\email{xinghu@zju.edu.cn}

\author{Xin Xia}
\orcid{0000-0002-6302-3256} 
\affiliation{    
  \institution{\zju}
  \city{Hangzhou}
  \country{China}
}
\affiliation{    
  \institution{Hangzhou High-Tech Zone (Binjiang) Institute of Blockchain and Data Security}
  \city{Hangzhou}
  \country{China}
}
\email{xin.xia@acm.org}

\begin{abstract}
Fuzzing is widely used for detecting bugs and vulnerabilities, with various techniques proposed to enhance its effectiveness.
To combine the advantages of multiple technologies, researchers proposed ensemble fuzzing, which integrates multiple base fuzzers. 
Despite promising results, state-of-the-art ensemble fuzzing techniques face limitations in resource scheduling and performance evaluation, leading to unnecessary resource waste.
In this paper, we propose \toolName, a novel ensemble fuzzing framework that dynamically schedules resources during the ensemble fuzzing campaign.
We designed a novel resource scheduling algorithm based on the upper confidence bound algorithm to reduce the resource consumption of ineffective base fuzzers.
Additionally, we introduce a multidimensional seed evaluation strategy, which considers multiple metrics to achieve more comprehensive fine-grained performance evaluation.
We implemented \toolName as a prototype tool and evaluated its effectiveness on Google's fuzzer-test-suite as well as real-world open-source projects. 
Results show that \toolName outperforms existing state-of-the-art base fuzzers and ensemble fuzzing techniques, detecting 20 bugs in real-world open-source projects—five previously unknown and three classified as CVEs. 
\end{abstract}

\begin{CCSXML}
<ccs2012>
   <concept>
       <concept_id>10011007.10011074.10011099.10011102.10011103</concept_id>
       <concept_desc>Software and its engineering~Software testing and debugging</concept_desc>
       <concept_significance>500</concept_significance>
       </concept>
   <concept>
       <concept_id>10002978.10003022.10003023</concept_id>
       <concept_desc>Security and privacy~Software security engineering</concept_desc>
       <concept_significance>500</concept_significance>
       </concept>
 </ccs2012>
\end{CCSXML}

\ccsdesc[500]{Software and its engineering~Software testing and debugging}
\ccsdesc[500]{Security and privacy~Software security engineering}

\keywords{
Ensemble Fuzzing,
Vulnerability Detection,
Resource Scheduling
}

\maketitle
\section{Introduction}

\label{sec_intro}
Fuzzing is one of the most widely used testing techniques for detecting bugs and vulnerabilities~\cite{Zhu_roadmap_survey, Manes_fuzzing_survey, CHEN_survey_2018118, BEAMAN_sruvey_2022102813, Poncelet_so_many}.
Recently, various types of fuzzers, including generation-based, mutation-based, and hybrid fuzzers, have been proposed.
However, existing studies~\cite{Zhu_roadmap_survey, Poncelet_so_many} show that no universal fuzzer consistently outperforms others.
Consequently, a natural idea to improve fuzzing effectiveness is to leverage multiple distinct base fuzzers simultaneously, which is known as \textit{ensemble fuzzing}.
Effective ensemble fuzzing requires scheduling the appropriate amount of resources (e.g., CPU cores or Docker containers) for each base fuzzer.
Most approaches~\cite{enfuzz, cupid} schedule a fixed amount of resources to each base fuzzer at startup, which remains unchanged throughout the ensemble fuzzing campaign.
Since the efficiency of each fuzzer is not always constant, the performance prediction before the actual fuzzing is inaccurate~\cite{cupid,enfuzz,fu2023autofz}, rendering fixed scheduling inefficient and inflexible.
The recently proposed approach Autofz~\cite{fu2023autofz} addresses this problem by splitting the fuzzing campaign into several rounds, each consisting of the Preparation Phase and the Focus Phase.
The Preparation Phase profiles the performance of all base fuzzers, while the Focus Phase performs continuous fuzzing based on allocated resources.
However, there are several issues to be addressed.
\ding{182} \textbf{Resource waste}. Autofz must execute all base fuzzers, including those with low performance, during the Preparation Phase to profile their performance.
This causes it to waste resources on inefficient fuzzers.
\ding{183} \textbf{Randomness-driven scheduling}. Autofz schedules resources solely based on performance rankings obtained in the Preparation Phase, ignoring the historical performance of base fuzzers. Consequently, its decisions are susceptible to the inherent randomness of fuzzing.
\ding{184} \textbf{Coarse performance metric}. Autofz utilizes a coarse-grained metric (the number of unique paths found) to evaluate performance, which lacks fairness and efficiency of resource scheduling and seed synchronization~\cite{invariants}.

\noindent {\bf Our Approach.} To address these problems, we propose a novel ensemble fuzzing framework, namely \textbf{\toolName}, which uses a light-overhead \textit{dynamic resource scheduling algorithm} with a \textit{multidimensional seed evaluation strategy}.
\toolName splits the ensemble fuzzing campaign into several rounds.
In each round, \toolName schedules each resource unit according to the performance of each base fuzzer and performs fuzzing, then evaluates the seeds generated by each base fuzzer and updates its performance accordingly.

\noindent {\bf Dynamic Resource Scheduling}. 
For Problem~\ding{182}, different from existing approaches~\cite{fu2023autofz}, \toolName achieves dynamic resource scheduling based on existing execution results, thereby getting rid of the dependence on the Preparation Phase and reducing the waste of resources on inefficient fuzzers.
Moreover, for problem~\ding{183}, \toolName formulates the dynamic resource scheduling problem as a special multi-armed bandit (MAB) problem ~\cite{MAB}, and introduces the upper confidence bound (UCB) algorithm, considering both the overall performance of the fuzzer and its exploration value.
In this paper, each base fuzzer is treated as a bandit, and scheduling a resource unit to a base fuzzer is analogous to pulling a bandit's arm once.
We schedule multiple resource units in a round, which is equivalent to pulling arms multiple times in one round.
Additionally, due to the randomness of fuzzing, the subsequent rewards of base fuzzers cannot be accurately predicted.
Therefore, \toolName modifies the existing algorithm and uses the pull count of the previous round instead of the cumulative rounds to expand the impact of exploration value on the results.
To optimize resource usage, we also introduce schedule fine-tuning midway through each round, reallocating resources from fuzzers that no longer generate beneficial seeds.

\noindent {\bf Multidimensional Seed Evaluation}.
For Problem~\ding{184}, after fuzzing based on the scheduled resources, \toolName comprehensively evaluates the benefits of their generated seeds using multiple metrics, including the increase of overall edge or path coverage, the number of crashes triggered, and the number of executed deep edges and less frequently covered paths.
To our knowledge, we are the first to combine all these fine-grained metrics for seed evaluation and apply them to ensemble fuzzing.
Additionally, \toolName employs the same strategy for \underline{seed synchronization} between base fuzzers.
During the ensemble fuzzing campaign, \toolName maintains a global seed pool and periodically synchronizes it with each base fuzzer's local pool.

To evaluate the effectiveness of \toolName, we compared \toolName with the state-of-the-practice ensemble fuzzing technique EnFuzz~\cite{enfuzz} and the state-of-the-art technique Autofz~\cite{fu2023autofz} using benchmark Google's fuzzer-test-suite~\cite{fuzzer-test-suite}.
For a fair comparison, we implemented variants of \toolName and AutoFz, named \toolNameA and AutoFz$^\alpha$, respectively, ensuring they use the same base fuzzers as EnFuzz: AFL~\cite{afl}, AFLFast~\cite{aflfast}, FairFuzz~\cite{fairfuzz}, LibFuzzer~\cite{libfuzzer}, Radamsa~\cite{radamsa}, and QSYM~\cite{qsym}.
The results show that \toolNameA consistently and effectively achieves better performance than EnFuzz, Autofz$^\alpha$, and all base fuzzers in terms of the number of paths executed, branches covered, and unique crashes detected.
Specifically, compared to EnFuzz and Autofz$^\alpha$, \toolNameA averagely covers 10.63\% and 5.62\% more branches, executes 6.44\% and 3.74\% more paths, and triggers 11.54\% and 5.45\% more unique crashes, respectively.   
Furthermore, to take advantage of the most state-of-the-art modern fuzzers, we also implemented a new variant, namely \toolNameB, of \toolName with six base fuzzers: AFL++~\cite{aflpp}, HonggFuzz~\cite{hfuzz}, LibFuzzer, Radamsa, Angora~\cite{angora}, and QSYM. 
On the nine newest real-world versions of open-source projects, \toolNameB successfully reported 20 bugs, five of which are previously unknown, and three of them are CVEs. 

Overall, this paper makes the following main contributions:
\begin{itemize}
	\item {\bf Dynamic Resource Scheduling.} 
    We propose a novel dynamic resource scheduling method, formulating the problem as a special MAB problem, using a novel UCB-based algorithm to take into account both historical rewards and current performance.
	\item {\bf Multidimensional Seed Evaluation.} We propose a novel seed evaluation strategy for ensemble fuzzing, which evaluates the profit of seeds with multiple metrics.
	\item {\bf Novel Framework}. We implement a novel ensemble fuzzing framework, \toolName, to evaluate the performance of all base fuzzers with multiple metrics and dynamically schedule resources for base fuzzers. The code
can be found on our website~\cite{legion}.
	\item {\bf Comprehensive Evaluation.} We evaluated the effectiveness of \toolName on Google's fuzzer-test-suite and real-world open-source projects. 
	Results show that \toolName outperforms existing state-of-the-art base fuzzers and ensemble fuzzing techniques. 
\end{itemize}

\section{Background and Motivation Example}
\subsection{Ensemble Fuzzing}

Ensemble fuzzing has gained popularity in recent years and is widely adopted in industry, such as Google's OSS-Fuzz platform~\cite{OSS-Fuzz} and Microsoft's OneFuzz platform~\cite{OneFuzz}.
Specifically, ensemble fuzzing uses multiple base fuzzers to fuzz a program simultaneously and schedules resources according to a certain strategy.
EnFuzz~\cite{enfuzz} is the first to popularize this concept and remains state-of-the-art, proposing a method for defining the diversity of base fuzzers and selecting a diverse set for ensemble fuzzing accordingly.
Cupid~\cite{cupid} improves the method by running each base fuzzer on a set of benchmark programs multiple times, comparing the branch coverage probability distributions, and selecting a base fuzzer set that maximizes the likelihood of achieving a high branch coverage.
CollabFuzz~\cite{collabfuzz} enhances EnFuzz by allowing users to express different seed scheduling strategies.
Autofz~\cite{fu2023autofz} splits the fuzzing campaign into several rounds and dynamically schedules resources based on the results of the Preparation Phase in each round.

In this paper, we propose a novel dynamic resource scheduling algorithm and a multidimensional evaluation strategy to address the resource waste problem present in existing techniques.

\subsection{Multi-Armed Bandit}
The Multi-Armed Bandit (MAB) problem involves a conflict between exploitation and exploration~\cite{MAB, whittle1980multi}.
In this problem, a player faces multiple bandits (or arms), each with an unknown reward distribution.
Specifically, suppose $K$ arms are given, and each arm yields a random reward when it is selected to pull. 
The reward distribution $\mu_i$ associated with each arm $i \in \{1,2,\ldots,K\}$ is unknown.
At each time step, one arm $i$ is selected to pull, and a reward $\gamma_i$ is received.
The objective is to maximize the accumulated rewards over $T$ rounds of experimentation~\cite{whittle1980multi}.
In the exploration phase, the player selects different arms to gather more information about their reward distributions.
In the exploitation phase, the player chooses the arms that provide higher rewards based on prior knowledge~\cite{bubeck2012regret}.
To solve this problem, various algorithms have been proposed, such as $\epsilon$-greedy algorithm~\cite{vermorel2005multi}, Thompson Sampling, and the Upper Confidence Bound (UCB) algorithm~\cite{ucb}. However, $\epsilon$-greedy algorithm relies on blind random exploration that wastes valuable CPU cycles on ineffective fuzzers, while Thompson Sampling requires assuming prior probability distributions that are difficult to define for fuzzing tasks.
Therefore, in this paper, we design our approach based on the UCB algorithm.
The UCB algorithm solves the MAB problem by computing a sum of two components for each arm (i.e., exploitation value and exploration value).
Let $\hat{\mu}_i$ be the historical reward of arm $i$, $t$ be the total number of pulls, and ${n_i}$ be the number of times arm $ i$ has been selected.
The traditional UCB algorithm assigns a score $UCB_i(t)$ to arm $i$ as follows:
\begin{equation}
\resizebox{0.3\linewidth}{!}{
$
\text{UCB}_i(t)  =  \hat{\mu}_i + c \times \sqrt{\frac{2 \ln t}{n_i}}
$
}
\end{equation}
where $c$ is a preset adjustment parameter.
Among them, $\hat{\mu}_i$ represents the currently known information, and $c \times \sqrt{\frac{2 \ln t}{n_i}}$ is the confidence upper limit (exploration value), which indicates the exploration of unknown information.

In this paper, we formulate dynamic resource scheduling as a special MAB problem where arms are pulled multiple times within a single round.
We use the UCB algorithm to address this MAB problem and modify it to adapt to the randomness of fuzzing.
In the UCB algorithm used by \toolName, the exploration value refers to the uncertainty term added to the reward. It quantifies the potential rewards of allocating resources to a fuzzer that has limited performance data. The exploration value favors fuzzers that have been allocated fewer resources.
Based on these, this mechanism prioritizes high-performing base fuzzers while guaranteeing that under-utilized fuzzers are periodically re-evaluated as their uncertainty terms grow over time.

\subsection{Motivation Example}
\label{motivating}

\begin{figure}
	\centering
 	\includegraphics[width=0.65\linewidth]{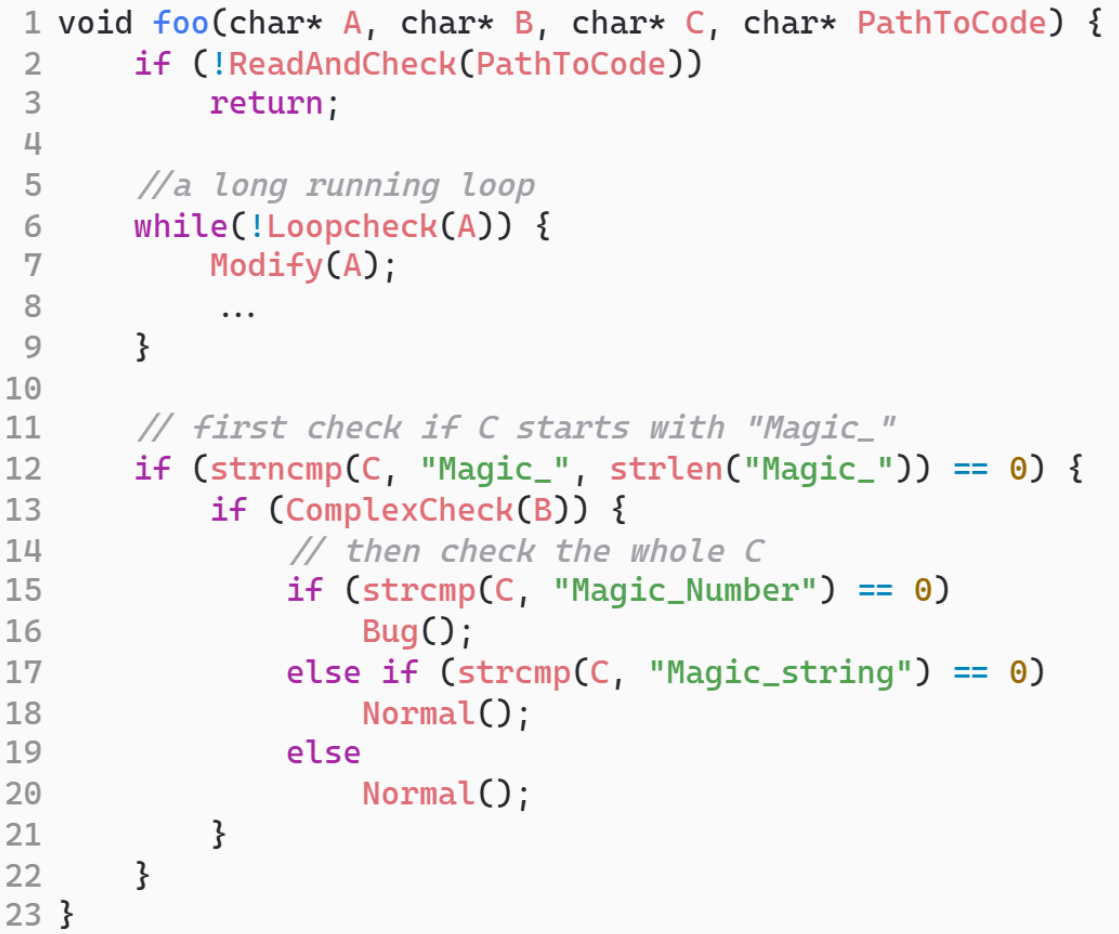}
    \caption{Motivation Example}\label{fig:motivating}
\end{figure}

Figure~\ref{fig:motivating} shows a synthetic motivation example in which the target function \texttt{foo} takes four strings as input:~\texttt{A}, \texttt{B}, \texttt{C}, and \texttt{PathToCode}.
First, the code file stored at \texttt{PathToCode} is checked against some constraints; if it violates them, \texttt{foo} terminates early (Lines 2--3).
Second, a long-running loop is executed, in which the input string \texttt{A} is repeatedly modified until it satisfies the constraint (Lines 6--9).
Next, \texttt{foo} checks if the input string \texttt{C} starts with \texttt{``Magic\_''} (Line 12), and then checks \texttt{B} against a complex constraint (Line 13).
Finally, \texttt{foo} checks if \texttt{C} is \texttt{``Magic\_Number''} or \texttt{``Magic\_String''} (Lines 15--20).
A bug can be triggered if \texttt{C} is \texttt{``Magic\_Number''} (Line 16). This simple example can pose various common challenges for fuzzing techniques, such as structured inputs~\cite{Olsthoorn_generating_structure}, long-running loops~\cite{pata}, complex constraints~\cite{qsym}, magic number checking~\cite{eclipser}. 
These challenges can either cause fuzzers to hang or prevent them from exploring relevant code~\cite{Zhu_roadmap_survey, Manes_fuzzing_survey}.

It is extremely hard for one fuzzing technique to address all of the above challenges.
For instance, some generation-based fuzzers such as TitanFuzz~\cite{titanfuzz} excel at generating code snippets, which can easily pass the first check of \texttt{foo} in Lines 2--3.
However, they can have difficulties in passing guards with complex constraints (e.g., in Line 6 and Line 13).
Some hybrid fuzzers such as QSYM~\cite{qsym} adopt heavy code analysis methods, e.g., concolic execution~\cite{concolic}, to solve complex constraints.
However, they are not likely to pass the code check in Line 2 as they cannot generate valid code snippets. 
Thus, there is a dire need to combine the strengths of different fuzzing techniques.
For example, a generation-based fuzzing technique can generate seeds passing the code check (Line 2) and be synchronized to a hybrid fuzzing technique to solve the remaining complex constraints (Line 6 and Line 13). 
Several techniques~\cite{enfuzz,cupid,fu2023autofz} have been proposed to combine different fuzzing techniques via an ensemble fuzzing campaign.
However, the effectiveness suffers from the inability to schedule the proper amount of resources for base fuzzers.

\noindent {\bf Dynamic Resource Scheduling.} 
During ensemble fuzzing, different amounts of resource units should be scheduled to different base fuzzers efficiently and dynamically based on the benefits generated by different base fuzzers.
For instance, resource units scheduled to base fuzzers specializing in code snippet generation should initially increase and later decrease once they have produced sufficient seeds, as determined by passing the check of \texttt{foo} in Lines 2–3.
It allows other fuzzers, such as hybrid ones, to receive more resource units at a later stage.
If the amount of resource units for each base fuzzer is fixed, it would not just hinder the maximum use of effective base fuzzers in different contexts, but also waste resources.
To achieve dynamic resource scheduling, the state-of-the-art technique Autofz~\cite{fu2023autofz} proposes to split the fuzzing campaign into several rounds and divide a round into two phases: the Preparation Phase and the Focus Phase. 
However, Autofz runs all base fuzzers in the Preparation Phase, including those inefficient ones, to evaluate the performance of base fuzzers and determine the resource scheduling for the Focus Phase.
For instance, generation-based fuzzers can help Autofz pass the check in Line 2.
After that, this fuzzer has no advantage over others, but still runs at each Preparation Phase.

\noindent {\bf Key Idea.} We propose a novel light-overhead algorithm to dynamically schedule resources during the ensemble fuzzing campaign based on the performance of each base fuzzer without a separate preparation phase.
For example, our algorithm can increase resources for base fuzzers specialized in tackling constraints in the long-running loop when few seeds pass the loop check in \texttt{foo} (Line 6), and decrease resources for them when sufficient seeds pass the loop check has been generated.
Section~\ref{sec:schedule} discusses the details of this algorithm.

\noindent {\bf Multidimensional Seed Evaluation.} 
When evaluating the benefits of seeds generated by each base fuzzer for resource scheduling and seed synchronization, existing ensemble fuzzing techniques use coarse-grained metrics.
Autofz~\cite{fu2023autofz} deems a seed beneficial only if it uniquely executes a program path.
Such strategies can lead to inefficient resource scheduling~\cite{invariants} and loss of beneficial seeds, which is observed more often for hidden deep program states. 
For instance, when running a base fuzzer $f$ specialized for passing string comparing checks, it may generate a seed assigning \texttt{C} with ``\texttt{Magic\_String}'' to pass the check in Line 12.
After that, Autofz switches to using the base fuzzer that solves complex constraints, and no resources are scheduled for $f$ regardless of whether line 13 is solved, since it is difficult for $f$ to cover more program paths (line 15) in the Preparation Phase (short running time).

\noindent {\bf Key Idea.} 
We propose a new seed evaluation strategy that comprehensively evaluates the benefits of seed from multiple dimensions, including whether seeds increase edge coverage or path coverage, cause any crashes, and cover deep edges and paths that are less frequently covered, which are presented in Section~\ref{sec:sync}.
For the above example of $f$, our strategy recognizes that the seeds generated by $f$ can cover less frequently covered edges (Line 12).
As a result, it allocates more resources to $f$, increasing its chances of generating seeds containing ``\texttt{Magic\_Number}''.
Moreover, all generated seeds are synchronized to other base fuzzers to increase the chances of directly triggering the bug after solving the constraint in Line 13.
Note that this specific code snippet is a synthetic example constructed to cleanly illustrate the challenges that fuzzing may face.
In practice, real-world program logic is highly complex, and these fuzzing challenges~\cite{eclipser,qsym,pata,Olsthoorn_generating_structure} frequently manifest in parallel.
This parallelism occurs as different seeds simultaneously explore different parts of the program, or as multiple distinct challenges compound within the same code segment.
Due to this complexity, relying on a single metric (e.g., standard branch coverage) is often insufficient to accurately assess a seed's progress or potential. This reality further motivates our multidimensional seed evaluation strategy.
\section{Our Approach: \toolName}\label{approach}
\subsection{Overview}

\begin{figure}[t]
	\centering
 \includegraphics[width=0.65\linewidth]{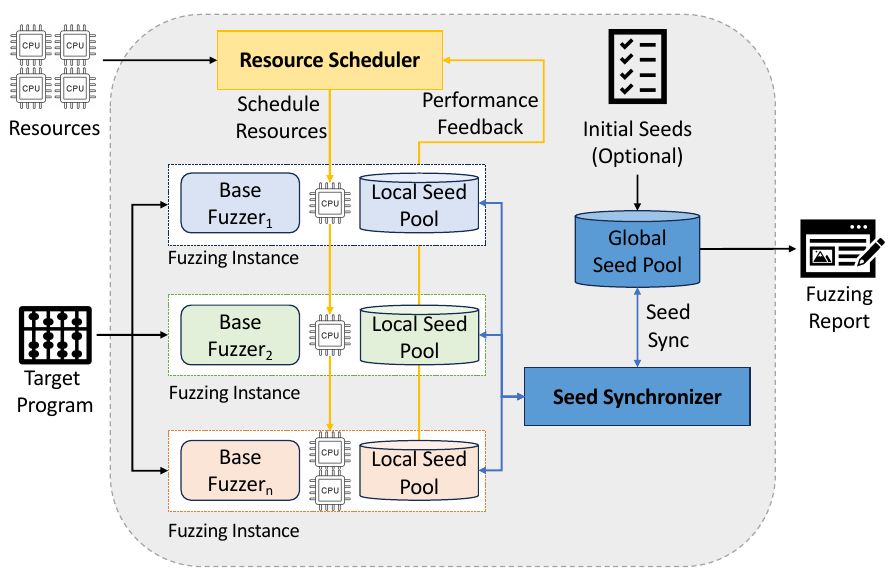}
	\caption{Overview of the \toolName Framework}\label{fig:overview}
\end{figure}

\begin{algorithm}[t]
	\small
	\Fn{\texttt{EnsembleFuzzing}$(P,R,S_{init}=\texttt{NULL})$}{
		$S_g	\leftarrow	S_{init}$\;
        \tcp*[h]{initialize fuzzing record} \\
		$M		\leftarrow	\texttt{init-record}(P)$\;
        \tcp*[h]{initialize evaluations for base fuzzers in the fuzzer pool $F$} \\
		$Q		\leftarrow	\{\tuple{f,\gamma=0,t=0} | f \in F\}$\; 
        \tcp*[h]{find deep edges} \\
        $D      \leftarrow  \texttt{find-deep}(P)$\;
        \tcp*[h]{initialize N} \\
        $N      \leftarrow  0 $\;
		\For{\rm each $i \in \{1,2,\ldots \texttt{FUZZ\_ROUND}\}$}{
            \tcp*[h]{schedule resource units for base fuzzers in Algorithm 2 and update N} \\
            $\phi \leftarrow  \texttt{schedule-resources}(F,Q,R,N) $\;
            \tcp*[h]{$\Delta=\{S_f\}$ are the corresponding final local seed pools} \\
            $\Delta\leftarrow   \texttt{fuzz-one-rd}(F,R,S_g,\phi,\texttt{ROUND\_TIME})$\;
			\tcp*[h]{update base fuzzer evaluation in Algorithm 2} \\
			$Q		\leftarrow	\texttt{perf-feedback}(P,F,M,D,Q,\Delta) $\; 
			\tcp*[h]{sync global seed pool with local pools} \\
			$\tuple{S_g,M}	\leftarrow	\texttt{seed-sync}(S_g,P,M,D,\Delta) $\; 
		}
		\Return{$\texttt{post-fuzz}(S_g)$}
	}
	\caption{Workflow of the \toolName Framework\label{alg_overview}}

\end{algorithm}

Figure~\ref{fig:overview} illustrates the overview of \toolName, which takes a target program $P$, a set of testing resource units $R=\{r_1,r_2,\ldots,r_t\}$, and optionally a set of initial seeds $S_{init}$ as input, 
and conducts an ensemble fuzzing campaign with a set of pre-selected base fuzzers $F$. 
A resource unit $r\in R$ represents the minimal unit to schedule for each base fuzzer to run.
Depending on the scheduling granularity, a resource unit can be a CPU core, a docker container, a server, etc.
Following existing ensemble fuzzing techniques~\cite{enfuzz,cupid,fu2023autofz}, we define a CPU core as a resource unit.
Note that our methodology has no restriction on the resource type.
\toolName relies on a \emph{Resource Scheduler} to dynamically schedule the given resource units during the ensemble fuzzing campaign. 
When running on a resource unit, each base fuzzer $f\in F$ maintains a local seed pool $S_f$ that periodically synchronizes with the global seed pool $S_g$ controlled by a \emph{Seed Synchronizer}. 

Algorithm~\ref{alg_overview} shows \toolName's workflow.
\toolName initializes the global seed pool $S_g$ (Line 2, $S_{init}=\texttt{NULL}$ if not given), 
and the global fuzzing record $M$ (Line 3) with the given initial seed $S_{init}$. 
Specifically, \toolName keeps track of the hit count of each edge $e\in P$ and the number of triggered crashes. 
Next, an initial evaluation tuple for each base fuzzer $f\in F$ is initialized (Line 4), indicating that $f$ has been deployed but no evaluation has been made.
\toolName utilizes lightweight static analysis to find all \emph{deep} edges in $P$ (Line 5).
We describe the collection and use of such deep-edge information in more detail in Section~\ref{sec:sync}.
After the initialization, \toolName iterates multiple rounds to conduct ensemble fuzzing (Lines 7--11).
At each round, \toolName first generates a resource schedule $\phi: R\to F$ based on the current performance evaluation $Q$ and the used resource unit count of all base fuzzers $N$ (Line 8).
% fuzzing
Next, each base fuzzer $f\in F$ conducts a traditional continuous fuzzing campaign with resource units scheduled according to $\phi$, seeds in the global seed pool $S_g$ as initial seeds, and \texttt{ROUND\_TIME} as the time limit (Line 9).
Note that during each fuzzing round, \toolName monitors the process and fine-tunes the resource schedule $\phi$.
After all base fuzzers finish running, their corresponding local seed pools $\Delta=\{S_f|f\in F\}$ are used as performance feedback to update the performance evaluation (Line 10), and profitable seeds are uploaded to $S_g$ (Line 11).
Finally, after the finish of the ensemble fuzzing, the post-fuzzing analysis is conducted on $S_g$ and a report is returned (Line 12).
We describe the resource scheduling algorithm in Section~\ref{sec:schedule}, and the seed evaluation and synchronization strategy in Section~\ref{sec:sync}.

\subsection{Dynamic Resource Scheduling}\label{sec:schedule}
To evaluate and predict the performance of each base fuzzer in the ensemble fuzzing campaign, and schedule resources for base fuzzers accordingly, we formulate the dynamic resource scheduling problem as a special MAB problem.
Each base fuzzer is viewed as an arm associated with an unknown reward distribution.
Selecting a base fuzzer to run on a resource unit for one round can be viewed as selecting the corresponding arm to pull, and the reward can be measured by analyzing the final local seed pool. 
Therefore, the evaluation of a base fuzzer $f$ is represented as a tuple $\tuple{f,\gamma,t}$, where $\gamma$ is the accumulated reward and $t$ is the pulling count of the previous round.
Note that in one round, scheduling a resource unit to a base fuzzer is formulated as pulling the arm once.
Specifically, \toolName sets the $t$ as the number of resource units scheduled for $f$ at the previous round.
Within a round, $t$ might be adjusted when our resource schedule fine-tuning mechanism reallocates resource units. 
Algorithm~\ref{alg_sch} shows the algorithm of \toolName for scheduling resources (Lines 1--19) and dynamically updating evaluations (Lines 20--29).

\begin{algorithm}[t]
	\small
    \Fn{\texttt{schedule-resources}$(F,Q,R,N)$}{
        $\phi   \leftarrow  \emptyset$\;
        \For{\rm each $r\in R$}{
            $f_r \leftarrow \texttt{assign}(F,Q,N)$\;
            $\phi   \leftarrow  \phi \cup \{\tuple{f_r,r}\}$\;
        }
        \Return{$\phi$}
    }
	\Fn{\texttt{assign}$(F,Q,N)$}{
		$U		\leftarrow	\emptyset$\;
		$N	\leftarrow	N + \sum_{\tuple{f,\gamma,t}\in Q}t$\;
		\For{\rm each $\tuple{f,\gamma,t} \in Q$}{
			\If{$t \ne 0$}{
                $q	\leftarrow	\frac{\gamma}{\gamma+t}$; \tcp*[h]{bounded efficiency score} \\
				$u	\leftarrow	\sqrt{\frac{2\log N}{t}}$; \tcp*[h]{upper confidence bound} \\
				$U	\leftarrow	U \cup \{\tuple{f,q,u}\}$\;
			}
			\Else{
				$U	\leftarrow	U \cup \{\tuple{f,q_{init},u_{init}}\}$\;
				$Q	\leftarrow	Q/\{\tuple{f,\gamma,t}\}$\;
			}
		}
		$f'	\leftarrow	\texttt{soft-max}(U)$\;
		\Return{$f'$}
	}
	\Fn{\texttt{perf-feedback}$(P,F,M,D,Q=\{\tuple{f,\gamma,t}|f\in F\},\Delta=\{S_f|f\in F\})$}{
		\For{\rm each $S_f \in \Delta$} {
                $\tuple{f,\gamma,t} \leftarrow \texttt{get-corresponding-tuple}(Q, f)$\;
			\If {$\tuple{f,\gamma,t}$ is NULL } {
                    $\tuple{f,\gamma,t} \leftarrow \tuple{f,0,0}$\;
				$Q \leftarrow	Q \cup \{\tuple{f,\gamma,t}\}$\;
			}
			$\gamma_{new}		\leftarrow	\texttt{evaluate}(P,M,D,S_{f})$\;
			$\gamma'	\leftarrow \gamma + \gamma_{new}$\;
			$t'	\leftarrow	\texttt{get-pull-count}(f)$\;
			$Q	\leftarrow	\texttt{replace}(Q, \tuple{f,\gamma,t},\tuple{f,\gamma',t'})$\;
		}
		\Return{$Q$}
	}
	\caption{Dynamic Resource Scheduling}\label{alg_sch}
\end{algorithm}

\noindent \textbf{Resource Scheduling.}
In \texttt{schedule-resources}, when a new round begins, \toolName generates a resource schedule $\phi$ (Lines 1--6). 
Specifically, \toolName iterates over each resource unit and schedules it to a base fuzzer (Lines 3--5). For each resource unit, the function \texttt{schedule-resources} calls the function \texttt{assign} to select a base fuzzer based on the performance evaluation (Lines 7--19).

In the function \texttt{assign}, \toolName first checks whether a base fuzzer $f$ has been deployed before (i.e., whether the corresponding $t \ne 0$) (Line 11).
If it has, \toolName calculates the bounded efficiency score of $f$ as $q=\frac{\gamma}{\gamma+t}$ (Line 12, as exploitation value), where $\gamma$ is the total accumulated reward and $t$ is the pull count from the previous round.
This formulation is designed to satisfy several properties for resource scheduling:
\textbf{1. Boundedness:} The formulation acts as a bounded non-linear mapping, projecting rewards to $[0, 1)$. This bounded range ensures numerical stability when combined with the Softmax function. Because accumulated rewards are often substantial, such excessively large exploitation values would lead to exponential divergence during subsequent Softmax calculations, causing the system to prematurely collapse into a ``winner-takes-all'' state.
\textbf{2. Monotonicity in accumulated reward:} Taking the partial derivative of this formulation with respect to $\gamma$ yields $\frac{\partial q}{\partial \gamma} = \frac{t}{(\gamma+t)^2} > 0$. This guarantees that higher accumulated rewards always increase the exploitation value.
\textbf{3. Penalization of recent resource usage:} Taking the partial derivative with respect to $t$ yields $\frac{\partial q}{\partial t} = \frac{-\gamma}{(\gamma+t)^2} < 0$. This property ensures that allocating more resources in the previous round reduces the current exploitation value, thereby discouraging the over-allocation of resources to a single fuzzer. 
\textbf{4. Saturation behavior:} As $\gamma \to \infty$, the score $q \to 1$. This aligns with the diminishing returns of fuzzing.
Existing research~\cite{bohme2020fuzzing} shows that in the later stages of the fuzzing lifecycle, fuzzers often require exponentially more resources to achieve the same results. 
As $q$ saturates, the scheduler seamlessly transitions into an uncertainty-driven rotation system dominated by the exploration value $u$.
The penalty for heavy recent resource consumption shifts to $u$ (where a larger $t$ from the previous round decreases the value).
This dynamically forces the system to rotate resources among all base fuzzers in later stages, preventing the campaign from getting trapped in local optima while still favoring fuzzers with high historical rewards.
The exploration value is calculated as $u=\sqrt{\frac{2 \log N}{t}}$ (Line 13). 
Furthermore, the randomness of fuzzing introduces noise into the scheduler (e.g., a random spike in coverage due to a lucky mutation can affect scheduler efficiency).
To mitigate the impact of noise, \toolName uses the accumulated reward instead of the single-round reward when calculating the exploitation value.
As fuzzing progresses, the impact of noise on the scheduler gradually decreases.
If $f$ has never been assigned any resource units, the initial $q_{init}$ and $u_{init}$ are used (Line 16), and $f$ is removed from the candidate list (Line 17).
We set $q_{init}$ to $1$ and $u_{init}$ to positive infinity so that all basic fuzzers are deployed in the initial priming round to evaluate their performance.
Removing from the candidate list ensures that each basic fuzzer is only assigned resources once.
With awards $q$ and upper confidence bounds $u$ of each base fuzzer, \toolName selects a base fuzzer to assign the resource unit in a soft-max manner (Line 18).
Specifically, \toolName calculates the probability of a base fuzzer $f\in F$ being selected as $Prob(f)=\frac{exp(q_f+u_f)}{\sum_{f'\in F}exp(q_{f'}+u_{f'})}$, and randomly selects one.
Note that in the initial priming round, \toolName selects all base fuzzers.
Softmax avoids the winner-takes-all scenario of deterministic strategies by maintaining a probability distribution and mitigating the potential noise introduced by the randomness of fuzzing.
It ensures that second-best fuzzers still receive some resources (i.e., Prob > 0).
Furthermore, fuzzing rewards are highly non-stationary, making them difficult to model using standard probability distributions. Compared to Thompson sampling, which relies on maintaining a posterior distribution of rewards (e.g., a Gaussian distribution), combining softmax with UCB-based scores provides a model-free probabilistic approach and further mitigates noise introduced by randomness through the exploration value.
Even if a fuzzer performs poorly due to noise, it still has a certain probability of being selected later due to its exploration value.

\noindent \textbf{Performance Evaluation Feedback.}
After the allocated resources are exhausted, the accumulated rewards and pull counts are updated (The function \texttt{perf-feedback} in Lines 20--30). 
First, we attempt to get the corresponding $\tuple{f,\gamma,t}$ from $Q$ using the function \texttt{get-corresponding-tuple} (Line 22).
This function searches $Q$ based on $f$ and returns the matched tuple if found; otherwise, it returns NULL.
Then, we check whether $\tuple{f,\gamma,t}$ is NULL (Line 23).
If so, we assign a value to $\tuple{f,\gamma,t}$ and add it into $Q$, which typically occurs during the initial priming phase (Line 24-25).
Next, seeds from the local seed pool $S_f$, associated with each $f$ assigned with resources in this round, are evaluated (Line 26), and the result of the evaluation $\gamma_{new}$ is added to the accumulated reward as a new reward $\gamma'$ (Line 27).
We elaborate in detail on how we evaluate the seeds with our fine-grained strategy in Section~\ref{sec:sync}.
Finally, we get the pull count of $f$ in this round, and update it to $t$ (Line 28).

\noindent \textbf{Resource Schedule Fine-tuning.}
During the fuzzing campaign of each round, \toolName monitors the execution of the selected base fuzzer $f$ on each resource unit $r\in R$, and periodically evaluates seeds generated by $f$ for every \texttt{MONITOR\_TIME} seconds with our multidimensional strategy after half of the \texttt{ROUND\_TIME}.
If $f$ fails to generate any beneficial seeds for the last \texttt{MONITOR\_TIME} seconds, \toolName stops $f$ and reschedules $r$ to another selected base fuzzer $f'$.
Specifically, we reschedule $r$ to the base fuzzer with the highest overall reward for the current round. 
Note that if all selected base fuzzers fail to generate beneficial seeds, the round terminates early.
The pull count of each base fuzzer is accordingly adjusted and reflected by \texttt{get-pull-count} in Line 28 in Algorithm~\ref{alg_sch}.
Note that we let each base fuzzer run at least half of \texttt{ROUND\_TIME} first because some base fuzzers, such as hybrid ones, may need some time to start generating beneficial seeds, especially in the late rounds.
Additionally, this fine-tuning mechanism is different from the Preparation Phase of Autofz.
The preparation phases execute all base fuzzers to profile and rank them before the actual resource allocation begins, regardless of their efficiency.
This blind execution introduces unavoidable upfront overhead in every scheduling round.
In contrast, \toolName's fine-tuning mechanism is a lightweight feedback mechanism.
It only monitors a subset of the base fuzzers that have been allocated resources.
If an active fuzzer stagnates and fails to generate new seeds within a preset time period, the fine-tuning mechanism reclaims its resources and reallocates them to the currently best-performing fuzzer.
Therefore, \toolName could use fine-tuning to avoid resource waste caused by fuzzing randomness.

\subsection{Multidimensional Seed Evaluation}\label{sec:sync}

We evaluate seeds $S$ with diverse fine-grained metrics.
Specifically, when executing the target program with $S_{f}$, \toolName records
(1) the total number of newly covered edges $c_0$, 
(2) the total number of newly covered execution paths $c_1$,
(3) the total number of triggered crashes  $c_2$,
(4) the total number of covered deep edges $c_3$, and 
(5) the total number of covered less frequently covered edges $c_4$.
These five metrics selected construct a comprehensive evaluation framework covering three dimensions: (1) Breadth ($c_0, c_1$): fundamental metrics for measuring state-space expansion; (2) Goal ($c_2$): ensuring vulnerability discovery is directly prioritized; and (3) Depth ($c_3, c_4$): serving as leading indicators to identify fuzzers penetrating complex logic when standard coverage plateaus.

In this paper, we define an edge as a transition between two basic blocks in the program's control flow graph, which is similar to the metric used in AFL~\cite{afl}.
This is measured at the assembly/IR level via compile-time instrumentation. An edge is considered covered if the transition is executed at least once. We define a path as a unique execution trace, distinguished by the sequence of executed edges. During the fuzzing process, Legion utilizes a fine-grained rolling hash (i.e., MurmurHash3) to track execution traces. This high sensitivity allows the scheduler to detect subtle progress (e.g., slight changes in loop iterations) and award resources to fuzzers that remain active, preventing premature termination.
We consider covering deep edges and less frequently covered edges as signs of benefits.
The rationale is that covering these two types of edges often indicates the examination of unexplored program states that cannot be distinguished by edge/path coverage, and thus may trigger additional crashes~\cite{stfuzzer}.
We adopt a lightweight static analysis to determine whether an edge is deep (Line 5 in Algorithm~\ref{alg_overview}).
Specifically, we construct a call graph of the target program $P$.
We consider edges of each function whose depth (i.e., the length of the shortest call path) is larger than $\rho d_\mu$ as deep edges, where $d_\mu$ is the average depth and $\rho=1.5$ is a factor commonly used for identifying outliers~\cite{outlier}.
Moreover, we consider edges whose coverage count is less than $\frac{1}{2}\mu$ as less frequently covered edges, where $\mu$ is the average coverage count of all covered edges, and $\frac{1}{2}$ is empirically tested and selected.
Furthermore, we include crash triggering as a core dimension in our evaluation. As demonstrated by specialized fuzzing modes (e.g., AFL's Crash Exploration), crashing inputs are not merely results, which also frequently traverse complex or fragile state spaces. Mutating these crash-triggering seeds often allows base fuzzers to uncover adjacent faulty logic or bypass immediate vulnerabilities to reach deeper code regions, making them highly valuable indicators of a fuzzer's exploratory potential.

We use all the above metrics to calculate the reward of a seed $s$ when scheduling resources (Line 26 in Algorithm~\ref{alg_sch}).
Specifically, the reward $\gamma_{new}$ is calculated as
$
\gamma_s(\theta)= \sum_{j=0}^{4} \theta_j c_j
$,
where $\theta$ is a weight vector. 
\toolName dynamically tunes the weight vector $\theta$ at each round to balance the importance of different types of measurements.
To achieve this, for each measurement type $c_j$ where $0 \le j \le 4$, we calculate the standard deviation $\sigma_j$ of the $c_j$ on all selected base fuzzers, and set $\theta_j = \frac{\sigma_j}{\sum_{i=0}^{4}\sigma_j}$.
Note that we assign $\theta_i=1$ for all $0 \le i \le 4$ for resource schedule fine-tuning since we only need to evaluate each seed qualitatively to determine whether a base fuzzer can generate any beneficial seed.
This dynamic weighting based on standard deviation prioritizes the discriminative power of different metrics.
Metrics exhibiting higher variance among fuzzers are assigned higher weights, ensuring the scheduler responds to the signals that better distinguish fuzzer performance, rather than being dominated by metrics with high absolute values but low deviation.
Additionally, \toolName aggregates multidimensional metrics using unnormalized incremental values $c_j$.
This unnormalized design is supported by our use of increments rather than cumulative totals.
While cumulative coverage totals diverge into the tens of thousands, creating an unbridgeable 
Finally, traditional cross-metric normalization (e.g., Min-Max scaling to $[0, 1]$) may introduce noise. For example, suppose Fuzzer A discovers 20 new paths and 4 crashes, while Fuzzer B discovers 19 new paths and 8 crashes. If we normalize these increments before aggregation, the path metric maps A to 1.0 and B to 0.0, while the crash metric maps A to 0.0 and B to 1.0. A simple sum yields 1.0 for both fuzzers, so the scheduler considers them identical. In reality, Fuzzer B sacrificed only 1 marginal path to discover 4 additional crashes, which is a significantly higher reward.

\toolName also utilizes the above metrics for seed synchronization.
Similar to resource fine-tuning, \toolName evaluates each seed qualitatively and uploads beneficial seeds to the global pool $S_g$.
Note that when scheduling resources for base fuzzers, a base fuzzer can be scheduled to run on multiple resource units simultaneously.
Many modern fuzzers such as AFL++~\cite{aflpp} ship with built-in mechanisms to synchronize seeds across multiple running instances and optimize their fuzzing strategies accordingly for better performance.
To take advantage of such mechanisms, at each round \toolName groups fuzzing instances with the same base fuzzers and allow them to conduct partial seed synchronization with each other.

%\vspace{-5pt}
\section{Evaluation}\label{evaluation}

We aim to answer the following research questions:

\begin{itemize}
	\item \textbf{RQ1}: How effective is \toolName compared to the state-of-the-art ensemble fuzzing techniques? 
	\item \textbf{RQ2}: How effective is \toolName improve the test performance over its base fuzzers?
    \item \textbf{RQ3}: How effective is \toolName with the state-of-the-art base fuzzers in detecting bugs in real-world projects?
	\item \textbf{RQ4 (Ablation Study)}: How beneficial are Legion's dynamic resource scheduling algorithm and multidimensional seed evaluation strategy?
\end{itemize}

\subsection{Subjects and Metrics}
\noindent {\bf Subjects.}
We included 33 subjects to evaluate \toolName.
To answer RQ1, RQ2, and RQ4, we used fuzzer-test-suite~\cite{fuzzer-test-suite}, a widely used benchmark, having 24 popular open-source real-world projects written in C or C++ as listed in Table~\ref{tab:enfuzz_google}.
We chose this benchmark due to its popularity and diversity, which includes projects with different sizes and functionalities.
While FuzzBench is a newer standard, we selected the fuzzer-test-suite because our primary baseline, EnFuzz, is strictly coupled with the hybrid fuzzer QSYM, which is explicitly unsupported by FuzzBench due to modern kernel compatibility issues.
To answer RQ3, we randomly collected the newest versions of nine real-world open-source projects as listed in Table~\ref{tab:bug_real}, from the ones used in existing studies~\cite{fu2023autofz,enfuzz,collabfuzz} with at least 100 GitHub stars.
Note that both the benchmark and the subjects we collected are real-world open-source projects.

\noindent {\bf Metrics.}
We used distinct metrics to answer our research questions. For RQ1, RQ2, and the ablation study (RQ4), we used three widely used quantitative metrics to evaluate performance: \textit{the number of covered branches}, \textit{the number of executed paths}, and \textit{the number of triggered unique crashes}.
We followed the same practice as Autofz~\cite{fu2023autofz} and used the top 3 frames of stack traces to de-duplicate unique crashes.
Moreover, for RQ3, which evaluates on real-world projects, the metrics are the number of real-world bugs (including CVEs and previously unknown vulnerabilities) discovered and the coverage.
For each identified bug, we searched the project’s issues to determine whether it was a CVE or previously unknown.

\subsection{Settings and Analysis Procedures}
\noindent {\bf RQ1 Analysis Procedure.} 
In RQ1, our goal is to evaluate the effectiveness of \toolName relative to other ensemble fuzzing frameworks, rather than the intrinsic capabilities of the base fuzzers.
Therefore, when evaluating the effectiveness of ensemble fuzzing, we do not include base fuzzers (e.g., AFL++) in the baseline but focus on comparisons with other ensemble fuzzing tools (i.e., EnFuzz~\cite{enfuzz} and Autofz~\cite{fu2023autofz}).
Because the implementation of our baseline, EnFuzz, is deeply coupled with six specific base fuzzers (i.e., AFL (Commit 6103710)~\cite{afl}, AFLFast (Commit d1d54ca)~\cite{aflfast}, FairFuzz (Commit e529c1f)~\cite{fairfuzz}, LibFuzzer (clang version 15.0.7)~\cite{libfuzzer}, Radamsa (Commit 5c32c29)~\cite{radamsa}, and QSYM (Commit 153a0b4)~\cite{qsym}), we strictly configured Autofz and \toolName (referred to as Autofz$^\alpha$ and \toolNameA) to use this same set. 
We use Doxygen (1.9.1-2ubuntu2)~\cite{doxygen} to construct the call graph of the target program statically. 

\underline{Settings.} Following EnFuzz and Autofz, a resource unit was set to a CPU core. 
For each subject in the Google fuzzer-test-suite, we ran \toolNameA, EnFuzz, and Autofz$^\alpha$ independently for 12 hours with six cores (72 CPU hours). 
We ran EnFuzz and Autofz$^\alpha$ with the same configurations in their evaluation.
Specifically, Autofz$^\alpha$ splits the fuzzing into several rounds.
Each round has 600 seconds, of which 300 seconds is for the Preparation Phase and 300 seconds is for the Focus Phase. 
In the Preparation Phase, it runs each base fuzzer for 30 seconds until the best one is found.
To provide a fair comparison with Autofz$^\alpha$, we accordingly set the same hyperparameters of \toolNameA.
Specifically, we set the time limit for each round as $\texttt{ROUND\_TIME}=600$ (accordingly $\texttt{FUZZ\_ROUND}=72$) and the time interval for fine-tuning as $\texttt{MONITOR\_TIME}=30$. 
The initial seeds for all experiments were the same.
We used the seeds provided by the fuzzer-test-suite or an empty seed if none was provided.
All experiments were repeated ten times.

\noindent {\bf RQ2 Analysis Procedure.} 
{In RQ2, our goal is to demonstrate that \toolName’s performance is not attributable to a particularly strong base fuzzer.}
Therefore, we further compared the performance of \toolNameA and all its base fuzzers on fuzzer-test-suite.
Same as the experiments to answer RQ1, for each subject, we ran each base fuzzer independently for 12 hours in parallel with six CPU cores and repeated ten times.
Note that we configured each base fuzzer in parallel mode (one master fuzzing process and five secondary fuzzing processes).

\noindent {\bf RQ3 Analysis Procedure.}
We implemented another variant of \toolName, namely \toolNameB, using the state-of-the-art fuzzers: AFL++ (4.06a)~\cite{aflpp}, HonggFuzz (Version 2.5)~\cite{hfuzz}, LibFuzzer (clang version 15.0.7)~\cite{libfuzzer}, Radamsa (Commit 5c32c29)~\cite{radamsa}, Angora (Version 1.3.0)~\cite{angora}, and QSYM (Commit 153a0b4)~\cite{qsym}.
AFL++, HonggFuzz, and LibFuzzer are the coverage-guided mutation-based fuzzers adopted by the popular fuzzing service OSS-Fuzz~\cite{OSS-Fuzz} hosted by Google. 
Radamsa is the representative generation-based fuzzer, and Angora and QSYM are representative hybrid fuzzers.
We let EnFuzz use the same base fuzzers as in RQ1 because its implementation is deeply coupled with base fuzzers, and configured Autofz to use the same six base fuzzers as \toolNameB, denoted as AutoFz$^\beta$.
We ran \toolNameB, \toolNameA, EnFuzz, Autofz$^\beta$, and each base fuzzer (parallel mode) on real-world projects, and compared their bug-detecting capabilities. 
All other settings are the same as RQ1.

\noindent {\bf RQ4 Analysis Procedure.} %To answer RQ3, 
We built four variants of \toolNameA for the ablation study: \toolNameC and  \toolNameE for analyzing the impact of the dynamic resource scheduling, \toolNameD for analyzing the impact of the multidimensional seed evaluation, and \toolNameF for analyzing the impact of the seed synchronization.
In each round of selecting the base fuzzer for the resource unit run, \toolNameC randomly selects the base fuzzer, and \toolNameE allocates resources evenly across each base fuzzer.
\toolNameD follows the same seed evaluation strategy as Autofz$^\alpha$, and is considered beneficial only if the seed can globally cover a unique program path.
This evaluation strategy is adopted by \toolNameD in resource scheduling and seed synchronization.
By comparing \toolName against \toolNameD, we evaluate our multidimensional strategy as a cohesive whole rather than statically ablating individual metrics. We do not conduct experiments that isolate and remove a single metric (e.g., only removing the crash metric) because \toolName's core contribution is the dynamic, standard deviation-based integration of these diverse signals.
Additionally, \toolNameF removes the seed synchronization, and each base fuzzer maintains its own seed pool.
A component is considered beneficial if its ablation results in a measurable drop in coverage and crash counts.
All settings in these experiments are the same as the ones in RQ1.
All variants use the same six base fuzzers as those of \toolNameA.

\noindent {\bf Environment.}
All experiments were conducted on a 64-bit machine with 24 cores (Intel(R) Core(TM) i9-14900K CPU @3.2GHz), 64GB of memory, and Ubuntu 16.04 LTS as the host OS which is the latest version of all tools in our evaluations can run~\cite{ubuntu}.
We do not set the performance mode or pin fuzzers to specific cores since both Legion and Autofz re-deploy fuzzers at each round.
The binary of each target program was hardened by AddressSanitizer~\cite{addresssanitizer} and UndefinedBehaviorSanitizer~\cite{ubsanitizer} to detect latent bugs.

\subsection{RQ1-Compared with Baselines}

\begin{table*}
	\centering
 \scriptsize
	\caption{RQ1. Comparison results between \toolNameA, EnFuzz, and Autofz$^\alpha$ on fuzzer-test-suite for ten times}\label{tab:enfuzz_google}
    %\vspace{-0.2cm}
	\begin{threeparttable}
		\setlength{\tabcolsep}{2pt}
		%\end{tabular}
		    \begin{tabular}{c|lll|lll|lll}
		    	\toprule
		    	\multicolumn{1}{l|}{\multirow{2}[0]{*}{\textbf{Project, abbr.}}} & \multicolumn{3}{c|}{\textbf{\# Branch}} & \multicolumn{3}{c|}{\textbf{\# Path}} & \multicolumn{3}{c}{\textbf{\# Crashes}} \\
		        & \textbf{\toolNameA} & \textbf{EnFuzz (P)} & \textbf{Autofz$^\alpha$ (P)} & \textbf{\toolNameA} & \textbf{EnFuzz(P)} & \textbf{Autofz$^\alpha$(P)} & \textbf{\toolNameA} & \textbf{EnFuzz} & \textbf{Autofz$^\alpha$} \\
		    	\midrule
		    	\rowsh \multicolumn{1}{l|}{boringssl, BS} & \textbf{4,155} & 3,581 (<0.01)  & 3,972 (<0.01)  & \textbf{7,529} & 7,028 (<0.01) & 7,385 (<0.01) & \textbf{1} & \textbf{1} & \textbf{1} \\
		    	 \multicolumn{1}{l|}{c-ares, CA} & \textbf{285} & \textbf{285} (0.98) & \textbf{285} (0.98) & \textbf{262} & 207 (<0.01)  & 251 (<0.01)   & \textbf{3} & \textbf{3} & \textbf{3} \\
		    	\rowsh \multicolumn{1}{l|}{guetzli, GU} & \textbf{3,679} & 2,199 (<0.01)  & 3,119 (<0.01)  & \textbf{4,498} & 4,013 (<0.01)  & 4,211 (<0.01)  & \textbf{1} & \textbf{1} & \textbf{1} \\
		    	\multicolumn{1}{l|}{lcms, LC} & \textbf{4,705} & 4,012 (<0.01)  & 4,328 (<0.01)  & \textbf{2,489} & 2,331 (<0.01)  & 2,487 (<0.01)  & \textbf{2} & \textbf{2} & \textbf{2} \\
		    	 \rowsh\multicolumn{1}{l|}{libarchive, LA} & \textbf{13,902} & 13,725 (<0.01)  & 13,652 (<0.01)  & \textbf{31,046} & 29,878 (<0.01)  & 30,874 (<0.01)  & \textbf{1} & \textbf{1} & \textbf{1} \\
		    	 \multicolumn{1}{l|}{libssh, LS} & \textbf{625} & 614 (<0.01)   & 614 (<0.01)   & \textbf{352} & 208 (<0.01)   & 286 (<0.01)   & \textbf{2} & \textbf{2} & \textbf{2} \\
		    	\rowsh \multicolumn{1}{l|}{libxml2, LX} & \textbf{21,744} & 19,314 (<0.01)  & 20,922 (<0.01)  & \textbf{35,819} & 32,485 (<0.01)  & 34,153 (<0.01)  & \textbf{3} & \textbf{3} & \textbf{3} \\
		    	 \multicolumn{1}{l|}{openssl-1.0.1, OS-0.1} & \textbf{4,685} & 4,399 (<0.01)  & 4,568 (<0.01)  & \textbf{4,498} & 4,380 (<0.01)  & 4,403  (<0.01) & \textbf{4} & 3     & \textbf{4} \\
		    	\rowsh  \multicolumn{1}{l|}{openssl-1.0.2, OS-0.2} & \textbf{4,258} & 4,039 (<0.01)  & 4,172 (<0.01)  & \textbf{4,903} & 4,657 (<0.01)  & 4,812 (<0.01)  & \textbf{6} & \textbf{6} & \textbf{6} \\
		    	 \multicolumn{1}{l|}{openssl-1.1.0, OS-1.0} & \textbf{9,039} & 8,425 (<0.01)  & 8,791 (<0.01)  & \textbf{4,738} & 4,529 (<0.01)  & 4,639 (<0.01)  & \textbf{6} & 5     & \textbf{6} \\
		    	\rowsh  \multicolumn{1}{l|}{pcre2, PC} & \textbf{54,735} & 50,135 (<0.01)  & 52,633 (<0.01)  & \textbf{87,192} & 85,093 (<0.01)  & 86,325 (<0.01)  & \textbf{7} & 6     & \textbf{7} \\
		    	\multicolumn{1}{l|}{proj4, PJ} & \textbf{915} & 805 (<0.01)   & 892 (<0.01)   & \textbf{697} & 652 (<0.01)   & 681 (0.53)  & \textbf{3} & \textbf{3} & 2  \\
		    	\rowsh \multicolumn{1}{l|}{re2, RE} & \textbf{18,751} & 16,335 (<0.01)  & 17,425 (<0.01)  & \textbf{17,165} & 16,893 (<0.01)  & 17,049 (<0.01)  & \textbf{1} & \textbf{1} & \textbf{1} \\
		    	\multicolumn{1}{l|}{woff2, WF} & \textbf{3,779} & 2,711 (<0.01)  & 3,289 (<0.01)  & \textbf{1,126} & 938 (<0.01)   & 985 (<0.01)   & \textbf{0} & \textbf{0} & \textbf{0} \\
		    	\rowsh \multicolumn{1}{l|}{freetype2, FT} & \textbf{60,233} & 57,415 (<0.01)  & 58,991 (<0.01)  & \textbf{30,191} & 26,647 (<0.01)  & 27,146 (<0.01)  & \textbf{0} & \textbf{0} & \textbf{0} \\
		    	\multicolumn{1}{l|}{harfbuzz, HB} & \textbf{40,293} & 34,551 (<0.01)  & 35,725 (<0.01)  & \textbf{18,335} & 15,970 (<0.01)  & 16,024 (<0.01)  & \textbf{1} & \textbf{1} & \textbf{1} \\
		    	\rowsh \multicolumn{1}{l|}{json, JS} & \textbf{7,128} & 6,925 (<0.01)  & 7,013 (<0.01)  & \textbf{1,247} & 1,138 (<0.01)  & 1,153 (<0.01)  & \textbf{3} & \textbf{3} & \textbf{3} \\
		    	\multicolumn{1}{l|}{libjpeg, LJ} & \textbf{18,625} & 13,235 (<0.01)  & 15,334 (<0.01)  & \textbf{2,936} & 2,431 (<0.01)  & 2,674 (<0.01)  & \textbf{0} & \textbf{0} & \textbf{0} \\
		    	\rowsh \multicolumn{1}{l|}{libpng, LP} & \textbf{6,025} & 5,043 (<0.01)  & 5,524 (<0.01)  & \textbf{766} & 705 (<0.01)   & 749 (<0.01)   & \textbf{1} & 0     & \textbf{1} \\
		    	\multicolumn{1}{l|}{llvm, LL} & \textbf{63,415} & 60,102 (<0.01)  & 61,291 (<0.01)  & \textbf{10,546} & 9,977 (<0.01)  & 10,428 (<0.01)  & \textbf{3} & 1     & 2  \\
		    	\rowsh \multicolumn{1}{l|}{openthread, OT} & \textbf{5,520} & 5,031 (<0.01)  & 5,473 (<0.01)  & \textbf{1,498} & 1,469 (<0.01)  & 1,481 (<0.01)  & \textbf{3} & \textbf{3} & \textbf{3} \\
		    	\multicolumn{1}{l|}{sqlite, SL} & \textbf{3,259} & 3,007 (<0.01)  & 2,918 (<0.01)  & \textbf{617} & 507 (<0.01)   & 491 (<0.01)   & \textbf{3} & \textbf{3} & \textbf{3} \\
		    	\rowsh  \multicolumn{1}{l|}{vorbis, VB} & \textbf{14,006} & 12,135 (<0.01)  & 13,497 (<0.01)  & \textbf{1,658} & 1,628 (<0.01)  & 1,649 (<0.01)  & \textbf{4} & \textbf{4} & 3  \\
		    	\multicolumn{1}{l|}{wpantund, WP} & \textbf{45,741} & 42,115 (<0.01)  & 43,302 (<0.01)  & \textbf{6,019} & 5,657 (<0.01)  & 5,832 (<0.01)  & \textbf{0} & \textbf{0} & \textbf{0} \\
		    	\midrule
		    	\multicolumn{1}{c|}{\textbf{total}} & \textbf{409,502} & 370,138   & 387,730  & \textbf{276,127} & 259,421   & 266,168   & \textbf{58} & 52    & 55  \\
		    	\multicolumn{1}{c|}{\textbf{Improvement (\increase)}} & -     & \textbf{10.63\%} & \textbf{5.62\%} & -     & \textbf{6.44\%} & \textbf{3.74\%} & -     & \textbf{11.54\%} & \textbf{5.45\%} \\
		    	\bottomrule
		    \end{tabular}%
        %Note: o.s.-1.0.x indicates the project openssl-1.0.x.
		
  %\begin{tablenotes}
	%		\small
     %       \item[1] In the \textbf{Project} column, o.s.-1.0.x indicates the project openssl-1.0.x.
		%	\item[1] In the \textbf{Impr.} row, each column presents the improvement percentage of \toolNameA over \toolNameC and \toolNameD.
	%	\end{tablenotes}
        
  %\david{can you add a up red arrow for the improvement number? also reduce the width of the table to be within left column width. libjped branch num is wrong}
	\end{threeparttable}
 % \vspace{-0.4cm}
\end{table*}

{\bf \toolNameA can outperform EnFuzz and Autofz$^\alpha$ in all metrics (Table~\ref{tab:enfuzz_google}).} 
On average, \toolNameA can cover 39,364 (i.e., 10.63\%) and 21,772 (i.e., 5.62\%) more branches, execute 16,706 (i.e., 6.44\%) and 9,959 (i.e., 3.74\%) more paths, and trigger 6 (11.54\%) and 3 (i.e., 5.45\%) more unique crashes in Google's fuzzer-test-suite than EnFuzz and Autofz$^\alpha$, respectively. 
With its dynamic resource scheduling algorithms, Autofz$^\alpha$ outperforms EnFuzz on almost all projects.
Nonetheless, \toolNameA achieves nearly twice the improvement of Autofz$^\alpha$ over EnFuzz in terms of covered branches and executed paths. 
Moreover, \toolNameA triggered 58 unique crashes in 20 out of 24 projects, including all crashes triggered by Autofz$^\alpha$ (55) and EnFuzz (52). 
We also followed common practice~\cite{fu2023autofz,collabfuzz,fairfuzz} and conducted the Mann-Whitney U Test~\cite{mcknight2010mann} between \toolNameA and the other two tools on the branch and path coverage.
On branch coverage, \toolNameA is statistically differentiated (i.e., p-value $<$ 0.01) from EnFuzz and Autofz$^\alpha$ on all except for c-ares, where all three tools cover the same number of branches (p-value = 0.98).
For executed paths, \toolNameA is statistically differentiated from EnFuzz on all and from Autofz$^\alpha$ on all but proj4 (p-value = 0.53).  

\begin{figure*}[t]
   \begin{minipage}{0.65\linewidth}
	\centering
	\includegraphics[width=\linewidth]{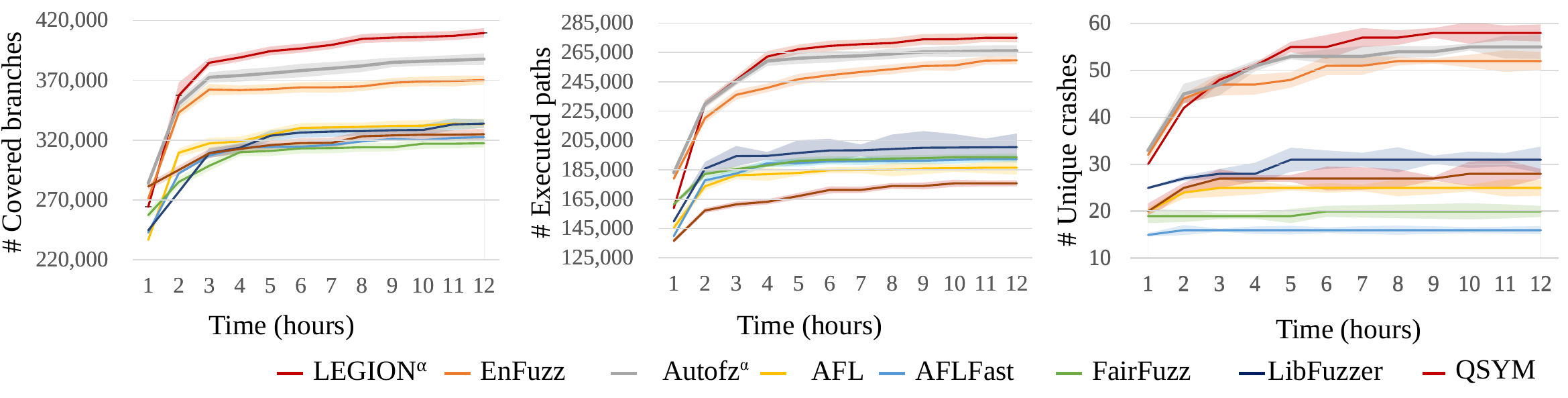}
	\caption{Average progressive results with 80\% confidence interval as shade on fuzzer-test-suite for ten times }
        \label{fig:progressive}
    \end{minipage} \hfill
    \begin{minipage}{0.3\linewidth}
            \centering
            \includegraphics[width=\linewidth]{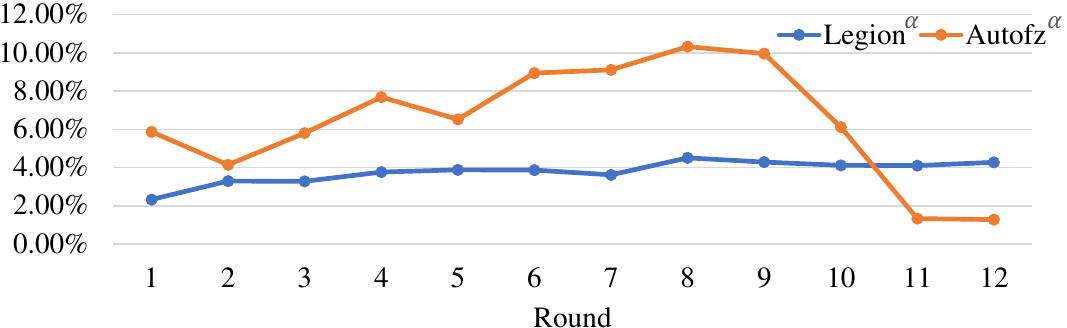}
            \caption{Overhead analysis for the first 12 rounds.}
            \label{fig:overhead}
    \end{minipage}
\end{figure*}

\begin{figure}[t]
	\centering
	\includegraphics[width=\linewidth]{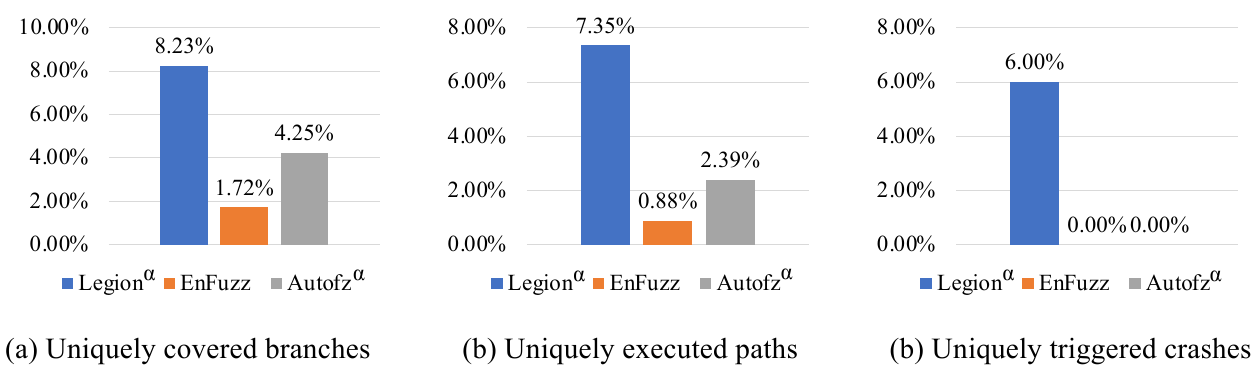}
	\caption{RQ1. Overlapping results between \toolNameA, EnFuzz, and Autofz$^\alpha$ 
	}
	\label{fig:overlapping-results}
\end{figure}

Figure~\ref{fig:progressive} shows the progressive average fuzzing results of \toolNameA, EnFuzz, and Autofz$^\alpha$ with 80\% confidence intervals. 
\toolNameA outperforms EnFuzz and Autofz$^\alpha$ in all three metrics after about three hours, and all tools gain marginally after six hours. 
Compared with EnFuzz, \toolNameA has lower gains in the first two hours, 
because \toolNameA spends additional execution time and resources on evaluating base fuzzers and scheduling resources after each round, 
which gain marginal benefits in early rounds due to the lack of sufficient data for analyses. 
The benefits of the scheduling algorithm quickly outweigh such additional cost, giving \toolNameA a substantial lead over EnFuzz after three hours. 
Compared with Autofz$^\alpha$, \toolNameA achieves slightly worse results in the first hour but quickly catches up.
While \toolNameA utilizes the fuzzing results of previous rounds for resource scheduling, Autofz$^\alpha$ conducts a Preparation Phase at each round.
Such a strategy allows Autofz$^\alpha$ to have sufficient data for analyses in the early rounds but also introduces a heavy overhead.
Figure~\ref{fig:overhead} shows the progressive overhead of \toolNameA and Autofz$^\alpha$ in the first two hours (i.e., 12 rounds).
For \toolNameA, the overhead comes from seed evaluation and resource scheduling for base fuzzers.
Therefore, we count the average percentage of CPU hours not used to run base fuzzers.
On the other hand, Autofz$^\alpha$ runs each base fuzzer during the Preparation Phase and selects a few to allocate resources in the Focus Phase.
Therefore, for Autofz$^\alpha$ we also count the CPU hours running base fuzzers that are \underline{not} scheduled with any resources in the Focus Phase. 
As Figure~\ref{fig:overhead} shows, in the early rounds, Autofz$^\alpha$ can quickly find the best base fuzzers and terminate the Preparation Phase early, yielding reasonable overhead.
However, when the coverage gain of each base fuzzer becomes marginal, it has to spend more time profiling, leading to more than 10\% overhead.
The overhead drops in the last two rounds when coverage begins to saturate, and Autofz$^\alpha$ schedules resources for all of them.
\toolNameA's overhead remains stable and reasonable while scheduling the proper amount of resources for base fuzzers to achieve better results in all three metrics.

\begin{table*}[]
	\centering
    %\small
    \scriptsize
	\caption{RQ2. Average number of branches and paths covered by \toolNameA and base fuzzers on fuzzer-test-suite for ten times}\label{tab:edge_google}
    \resizebox{\textwidth}{!}{
	\begin{threeparttable}
    \setlength{\tabcolsep}{2.5pt}
    \renewcommand{\arraystretch}{1.1}
    \begin{tabular}{l|lllllll|lllllll}
    \toprule
	\multirow{2}{*}{\textbf{Project}} & \multicolumn{7}{c|}{\textbf{\# Branches}}             & \multicolumn{7}{c}{\textbf{\# Paths}} \\
	& \multicolumn{1}{l}{\textbf{\toolNameAshort}} & \multicolumn{1}{l}{\textbf{AFL}} & \multicolumn{1}{l}{\textbf{AF.}} & \multicolumn{1}{l}{\textbf{FF.}} & \multicolumn{1}{l}{\textbf{Lib.}} & \multicolumn{1}{l}{\textbf{Rad.}} & \multicolumn{1}{l|}{\textbf{QSYM}} & \multicolumn{1}{l}{\textbf{\toolNameAshort}} & \multicolumn{1}{l}{\textbf{AFL}} & \multicolumn{1}{l}{\textbf{AF.}} & \multicolumn{1}{l}{\textbf{FF.}} & \multicolumn{1}{l}{\textbf{Lib.}} & \multicolumn{1}{l}{\textbf{Rad.}} & \multicolumn{1}{l}{\textbf{QSYM}} \\
	\midrule
	\rowsh \textbf{BS} & \textbf{4,155} & 3,365  & 3,194  & 3,523  & 3,419  & 3,296  & 3,178  & \textbf{7,529} & 3,197  & 2,733  & 3,879  & 5,623  & 3,271  & 2,777  \\
	\textbf{CA} & \textbf{285} & \bf 285   &  \bf 285   &  \bf 285   & 267   & 279   &  \bf 285   & \textbf{262} & 129   & 115   & 132   & 114   & 135   & 133  \\
	\rowsh \textbf{GU} & \textbf{3,679} & 2,514  & 2,208  & 1,135  & 2,493  & 3,019  & 2,855  & \textbf{4,498} & 3,375  & 3,471  & 3,392  & 3,694  & 3,815  & 3,317  \\
	\textbf{LC} & \textbf{4,705} & 3,517  & 3,319  & 3,524  & 2,897  & 3,614  & 3,655  & \textbf{2,489} & 1,571  & 1,662  & 1,582  & 1,435  & 1,668  & 1,585  \\
	\rowsh \textbf{LA} & \textbf{13,902} & 11,369  & 10,105  & 9,134  & 9,007  & 12,035  & 13,361  & \textbf{31,046} & 11,983  & 12,379  & 12,852  & 12,093  & 10,994  & 10,874  \\
	\textbf{LS} & \textbf{625} & 614   & 613   & 614   & 605   & 614   & 622   & \textbf{352} & 135   & 102   & 110   & 251   & 109   & 191  \\
	\rowsh \textbf{LX} & \textbf{21,744} & 13,812  & 12,019  & 10,054  & 11,259  & 16,542  & 18,852  & \textbf{35,819} & 14,241  & 15,735  & 16,833  & 15,012  & 13,998  & 12,083  \\
	\textbf{OS-0.1} & \textbf{4,685} & 4,135  & 4,002  & 3,918  & 3,996  & 4,125  & 4,415  & \textbf{4,498} & 3,812  & 4,053  & 4,115  & 3,769  & 4,022  & 3,735  \\
	\rowsh \textbf{OS-0.2} & \textbf{4,258} & 3,829  & 3,715  & 3,326  & 3,735  & 3,913  & 4,002  & \textbf{4,903} & 3,865  & 3,947  & 3,618  & 4,025  & 3,970  & 3,911  \\
	\textbf{OS-0.3} & \textbf{9,039} & 8,023  & 7,981  & 7,835  & 8,315  & 7,835  & 7,933  & \textbf{4,738} & 4,037  & 4,735  & 4,023  & 3,922  & 4,157  & 3,918  \\
	\rowsh \textbf{PC} & \textbf{54,735} & 51,389  & 50,113  & 49,836  & 48,315  & 52,923  & 50,113  & \textbf{87,192} & 78,136  & 79,053  & 77,106  & 80,253  & 78,209  & 73,421  \\
	\textbf{PJ} & \textbf{915} & 253   & 267   & 267   & 815   & 267   & 267   & \textbf{697} & 307   & 339   & 325   & 401   & 338   & 319  \\
	\rowsh \textbf{RE} & \textbf{18,751} & 11,319  & 12,416  & 13,872  & 15,033  & 15,979  & 16,532  & \textbf{17,165} & 11,973  & 12,389  & 13,003  & 11,254  & 12,034  & 10,375  \\
	\textbf{WF} & \textbf{3,779} & 120   & 120   & 120   & 2,693  & 120   & 120   & \textbf{1,126} & 102   & 113   & 135   & 148   & 97    & 148  \\
	\rowsh \textbf{FT} & \textbf{60,233} & 52,039  & 50,035  & 51,259  & 58,213  & 51,092  & 49,320  & \textbf{30,191} & 19,397  & 20,020  & 19,765  & 24,397  & 18,883  & 18,053  \\
	\textbf{HB} & \textbf{40,293} & 33,265  & 32,193  & 33,059  & 33,386  & 32,503  & 29,312  & \textbf{18,335} & 12,203  & 12,335  & 13,512  & 13,351  & 1,205  & 13,512  \\
	\rowsh \textbf{JS} & \textbf{7,128} & 6,015  & 5,135  & 5,524  & 6,623  & 6,014  & 5,992  & \textbf{1,247} & 870   & 913   & 885   & 1,125  & 903   & 768  \\
	\textbf{LJ} & \textbf{18,625} & 12,290  & 12,035  & 11,289  & 11,315  & 11,019  & 11,496  & \textbf{2,936} & 1,746  & 1,825  & 1,806  & 1,912  & 1,703  & 1,663  \\
	\rowsh \textbf{LP} & \textbf{6,025} & 4,283  & 4,439  & 4,013  & 5,125  & 4,005  & 4,118  & \textbf{766} & 603   & 601   & 512   & 637   & 591   & 588  \\
	\textbf{LL} & \textbf{63,415} & 58,125  & 57,369  & 53,963  & 58,218  & 56,215  & 52,135  & \textbf{10,546} & 8,023  & 8,819  & 8,654  & 9,039  & 8,251  & 7,653  \\
	\rowsh \textbf{OT} & \textbf{5,520} & 3,016  & 3,155  & 2,983  & 4,026  & 4,416  & 3,235  & \textbf{1,498} & 291   & 228   & 356   & 1,048  & 225   & 281  \\
	\textbf{SL} & \textbf{3,259} & 2,719  & 2,635  & 2,599  & 2,970  & 2,835  & 2,514  & \textbf{617} & 305   & 332   & 294   & 460   & 338   & 351  \\
	\rowsh \textbf{VB} & \textbf{14,006} & 11,153  & 10,988  & 12,576  & 11,026  & 12,346  & 9,974  & \textbf{1,658} & 1,329  & 1,488  & 1,572  & 1,191  & 1,425  & 1,351  \\
	\textbf{WP} & \textbf{45,741} & 41,325  & 40,886  & 41,026  & 40,075  & 39,914  & 41,235  & \textbf{6,019} & 4,833  & 4,936  & 5,154  & 5,332  & 4,921  & 4,913  \\
	\midrule
	\textbf{Total} & \textbf{409,502} & 338,774  & 329,227  & 325,734  & 343,826  & 344,920  & 335,521  & \textbf{276,127} & 186,463  & 192,323  & 193,615  & 200,486  & 175,262  & 175,920  \\
	\bf \increase & -     & \textbf{20.88\%} & \textbf{24.38\%} & \textbf{25.72\%} & \textbf{19.10\%} & \textbf{18.72\%} & \textbf{22.05\%} & -     & \textbf{48.09\%} & \textbf{43.57\%} & \textbf{42.62\%} & \textbf{37.73\%} & \textbf{57.55\%} & \textbf{56.96\%} \\
	\bottomrule
\end{tabular}%
         Note: \toolNameAshort = \toolNameA, AF. = AFLFast, FF. = FairFuzz, Lib. = LibFuzzer, and Rad. = Radamsa. 
		
	\end{threeparttable}
    }
\end{table*} 

\begin{table}[t]
	\centering
	\scriptsize
	\caption{RQ2. Average number of unique crashes triggered by \toolNameA and base fuzzers on fuzzer-test-suite  for ten times}\label{tab:crash_google}
	\begin{threeparttable}
		\begin{tabular}{l|p{0.05\columnwidth}p{0.08\columnwidth}p{0.08\columnwidth}p{0.08\columnwidth}p{0.07\columnwidth}p{0.07\columnwidth}p{0.09\columnwidth}}%
			\toprule
			\bf Project & \bf \toolNameAshort & \bf AFL & \bf AF. & \bf FF. & \bf Lib. & \bf Rad. & \bf QSYM \\
			\midrule
			\rowsh \bf BS & \bf 1&0&0&0&\bf1&0&0\\
			\bf CA& \bf 3&2&1&2&1&2&1\\
			\rowsh \bf GU&\bf1&0&0&0&\bf1&0&0\\
			\bf LC&\bf 2&1&0&1&\bf2&0&0\\
			\rowsh \bf LA&\bf 1&0&0&0&\bf 1&0&0\\
			\bf LS&\bf 2&0&0&0&1&0&1\\
			\rowsh \bf LX&\bf 3&1&1&1&2&\bf 3&1\\
			\bf OS-0.1& \bf 4&2&1&2&0&2&2\\
			\rowsh \bf OS-0.2&\bf 6&4&2&3&1&4&3\\
			\bf OS-1.0&\bf 6&4&3&3&2&4&5\\
			\rowsh \bf PC& \bf 7&5&3&2&5&5&4\\
			\bf PJ&\bf 3&1&0&1&0&2&1\\
			\rowsh \bf RE&\bf 1&\bf 1&0&0&\bf 1&0&\bf 1\\
			\bf WF&0&0&0&0&\bf 2&1&1\\
			\rowsh \bf FT&\bf 0&\bf 0&\bf 0&\bf 0&\bf 0&\bf 0&\bf 0\\
			\bf HB&\bf 1&0&0&\bf 1&\bf 1&0&\bf 1\\
			\rowsh \bf JS& \bf 3&1&0&0&1&\bf 3&2\\
			\bf LJ&\bf 0&\bf 0&\bf 0&\bf 0&\bf 0&\bf 0&\bf 0\\
			\rowsh \bf LP& \bf 1&0&0&0&0&0&0\\
			\bf LL& \bf 3&1&1&2&2&2&1\\
			\rowsh \bf OT&\bf 3&0&0&0&2&0&0\\
			\bf SL& \bf 3&0&0&0&2&1&1\\
			\rowsh \bf VB& \bf 4&2&4&2&3&1&3\\
			\bf WP&\bf 0&\bf 0&\bf 0&\bf 0&\bf 0&\bf 0&\bf 0\\
			\midrule
			\rowsh \bf Total& \bf 58&25&16&20&31&30&28\\
			\bf \increase&-&\bf 132.00\%&\bf 262.50\%&\bf 190.00\%&\bf 87.10\%&\bf 93.33\%& \bf 107.14\%\\
			\bottomrule
			
		\end{tabular}
        Note: AF. = AFLFast, FF. = FairFuzz, Lib. = LibFuzzer, and Rad. = Radamsa.
	\end{threeparttable}
\end{table}

Figure~\ref{fig:overlapping-results} displays the overlapping analysis results on the branches, paths, and unique crashes covered or triggered by \toolNameA, EnFuzz, and Autofz$^\alpha$. 
To prevent the influence of randomness on the results, we only analyzed those covered or triggered by a tool \emph{in all ten independent runs}.
Enfuzz uniquely covers only 1.72\% of branches and 0.88\% of paths.
\toolNameA uniquely covers 8.23\% of branches, and 7.35\% of paths, while Autofz$^\alpha$ only covers 4.25\% and 2.39\% ones, respectively.
Although EnFuzz and Autofz$^\alpha$ cover some unique branches and paths, \toolNameA triggers all unique crashes.

\greyboxb{Answer to RQ1:} {
\toolNameA outperforms EnFuzz and Autofz$^\alpha$ in all metrics (the number of branches and paths covered, and unique crashes).  \toolNameA achieves an average increase of 10.63\% and 5.62\% for branch explores more execution paths (6.44\% and 3.74\%), while triggering more unique crashes.}

\subsection{RQ2-Compared with Base Fuzzers}

{\bf On average, \toolNameA can outperform all of its base fuzzers in all metrics} (in Table~\ref{tab:edge_google} and \ref{tab:crash_google}).
Table~\ref{tab:edge_google} shows that on average, \toolNameA covers more branches (i.e., ranging from 18.72\% to 25.72\%) and executes more paths (i.e., ranging from 37.73\% to 57.55\%) than all of its base fuzzers. 
Moreover, Table~\ref{tab:crash_google} shows that \toolNameA outperforms all of its base fuzzers on the number of triggered unique crashes (e.g., 132.00\% more crashes than AFL), and 87.10\% more crashes than the best-performing base fuzzer, LibFuzzer.
Furthermore, as Figure~\ref{fig:progressive} shows, \toolNameA quickly outperforms all base fuzzers in all three metrics with a substantial lead. While LibFuzzer and QSYM alone gain similar branch coverage as other base fuzzers on libjpeg, \toolNameA combines their strengths and achieves much higher branch coverage than LibFuzzer and QSYM individually (i.e., improved by 50\%+).

Our empirical results in Table~\ref{tab:edge_google} and \ref{tab:crash_google} show that the performance of base fuzzers indeed varies on different projects.
For instance, although Radamsa achieves the best average branch coverage, it only achieves the best branch coverage in eight out of 24 projects.
Similarly, QSYM achieves the best branch coverage in eight projects but performs the worst for several projects such as freetype2 and vorbis.
As a comparison, by combining the strengths of all base fuzzers, \toolNameA consistently outperforms all base fuzzers on almost all projects, demonstrating the effectiveness of ensemble fuzzing.

\greyboxb{Answer to RQ2:} {
\toolNameA achieves higher branch and path coverage compared to using only a single base fuzzer by dynamically combining the advantages of these base fuzzers, and triggers 87.10\% more unique crashes than the best-performing single base fuzzer.}

\subsection{RQ3-Bug Detection in Real-world Projects}
{\bf Our empirical results in Table~\ref{tab:bug_real} show that \toolName triggers more bugs than EnFuzz, Autofz$^\beta$, and base fuzzers.}
%the unique bug detection results of selected fuzzing techniques.
\toolNameB and \toolNameA detect 20 and 13 unique bugs, respectively. EnFuzz and Autofz$^\beta$ detect 9 and 17 unique bugs, respectively, and all of them can be detected by \toolNameB. 
With the advantage of the state-of-the-art base fuzzers, \toolNameB gains 122.22\% and 53.85\% improvement over EnFuzz and \toolNameA, respectively.
Moreover, with the same base fuzzers, \toolNameB successfully detects 17.65\% more bugs than Autofz$^\beta$. 
We found that among the 20 bugs detected by \toolNameB, three of them are known CVEs (CVE-2022-4904, CVE-2023-28484, and CVE-2018-11097), and five of them are previously unknown (e.g., Integer Overflow in xpack).
Although \toolNameB discovers only one additional previously unknown bug compared to the strongest baseline, the evaluated projects have been extensively tested using existing fuzzing techniques.
Therefore, under a limited budget, the number of additional previously unknown bugs that can be exposed is constrained.
As a result, the number of unknown bugs alone may not fully capture the effectiveness of approaches.
% In terms of bug detection capability, \toolNameB identifies 20 unique bugs, outperforming all evaluated ensemble fuzzing.
We compared \toolNameB with its SOTA base fuzzers. Table~\ref{tab:bug_real} shows that without ensemble fuzzing, the best-performing base fuzzer AFL++ detects 15 unique bugs, while \toolNameB detects 20, including all the bugs detected by AFL++, achieving 33.33\% improvement.

\begin{figure}[t]
    \centering
	\includegraphics[width=0.75\linewidth]{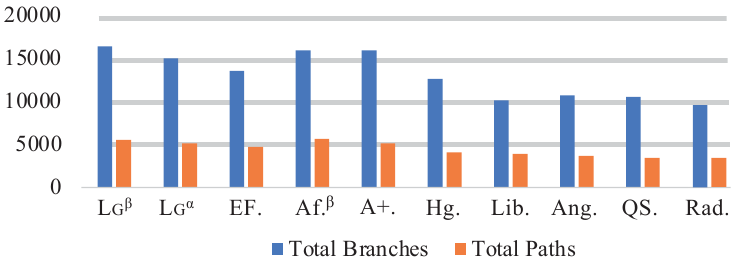}
    \caption{RQ3. Coverage results of tools on real-world projects
    }
    \label{fig:coverage_real}
\end{figure}

\begin{figure}[t]
    \centering
	\includegraphics[width=0.8\columnwidth]{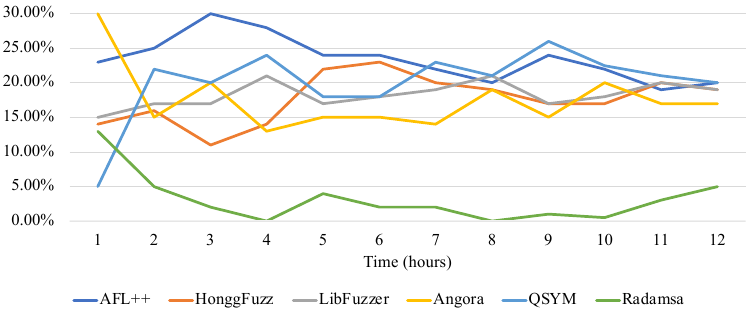}
    \caption{RQ3. Resource scheduling results for base fuzzers of \toolNameB}\label{fig:fuzzer_time_real}
\end{figure}

\begin{table}[t]
	\centering
\scriptsize
	\caption{RQ3. Bug detection results on real-world projects
 %\david{can we shrink it to one column width?}
 \vspace{-0.1cm}
}\label{tab:bug_real}
    \begin{threeparttable}
	%\begin{tabular}{l|c|cccccc}%{p{0.1\columnwidth}|p{0.08\columnwidth}p{0.08\columnwidth}p{0.09\columnwidth}|p{0.06\columnwidth}p{0.07\columnwidth}p{0.03\columnwidth}p{0.05\columnwidth}p{0.035\columnwidth}}
	%		\toprule
	%		\bf Project &\bf \toolNameBshort  &  \bf A+. & \bf Hg. & \bf Lib. & \bf Ang. & \bf QSYM & \bf Rad.   \\
     %       \midrule
	%		\rowsh c-ares &\bf 1&\bf 1&\bf 1 &\bf 1&\bf 1&\bf 1&0\\
	%		ffjpeg&\bf 2&\bf 2&0 &\bf 2&\bf 2&1&1\\
	%		\rowsh openjpeg&\bf 2&\bf 2&1&1& \bf 2& 0&1\\
	%		libxml2&\bf 5&4&3&3&3&2&3\\
	%		\rowsh xml2json&\bf 2&0&1&0&0&1&0\\
     %           xpack   &\bf 1&0&0&0&0&0&\bf 1\\
      %          \rowsh cstring &\bf 1&\bf 1 &\bf 1 &\bf 1 &\bf 1 &\bf 1 & 0 \\
       %         tinyexr &\bf 1 & \bf 1 & 0 & 0 & 0 & 0 & 0\\ 
        %    \rowsh jhead & \bf 5 & 3 & 2 & 3 & 4 & 4 & 1 \\
		%	\midrule
		%	\bf Total&20&14&9&11&13&10&7\\
		%	\bottomrule			
		%\end{tabular}
  %Note: \toolNameBshort = \toolNameB, A+ = AFL++, Hg. = HonggFuzz, Lib. = Libfuzzer, Ang. = Angora, and Rad. = Radamsa.% represents HonggFuzz, Libfuzzer, and Radamsa, respectively.
      \begin{tabular}
      %{l|cccc|cccccc}
    {p{0.15\columnwidth}|p{0.03\columnwidth}p{0.03\columnwidth}p{0.03\columnwidth}p{0.03\columnwidth}|p{0.03\columnwidth}p{0.03\columnwidth}p{0.03\columnwidth}p{0.03\columnwidth}p{0.04\columnwidth}p{0.03\columnwidth}}
      	\toprule
  	\textbf{Project} & \textbf{\toolNameBshort} & \textbf{\toolNameAshort} & \textbf{EF.} & \textbf{Af.$^\beta$} & \textbf{A+.} & \textbf{Hg.} & \textbf{Lib.} & \textbf{Ang.} & \textbf{QSYM} & \textbf{Rad.} \\
  	\midrule
  	\rowsh c-ares & \textbf{1} & \textbf{1} & \textbf{1} & \textbf{1} & \textbf{1} & \textbf{1} & \textbf{1} & \textbf{1} & \textbf{1} & 0 \\
  	ffjpeg & \textbf{2} & \textbf{2} & \textbf{2} & \textbf{2} & \textbf{2} & 0     & \textbf{2} & \textbf{2} & 1     & 1 \\
  		\rowsh openjpeg & \textbf{2} & 1     & 1     & \textbf{2} & \textbf{2} & 1     & 1     & \textbf{2} & 0     & 1 \\
  	libxml2 & \textbf{5} & 4     & 2     & 4     & \textbf{5} & 3     & 3     & 3     & 2     & 3 \\
  		\rowsh xml2json & \textbf{2} & 1     & 0     & \textbf{2} & 0     & 1     & 0     & 0     & 1     & 0 \\
  	xpack & \textbf{1} & 0     & 0     & \textbf{1} & 0     & 0     & 0     & 0     & 0     & \textbf{1} \\
  		\rowsh cstring & \textbf{1} & \textbf{1} & \textbf{1} & 0     & \textbf{1} & \textbf{1} & \textbf{1} & \textbf{1} & \textbf{1} & 0 \\
  	tinyexr & \textbf{1} & 0     & 0     & \textbf{1} & \textbf{1} & 0     & 0     & 0     & 0     & 0 \\
  		\rowsh jhead & \textbf{5} & 3     & 2     & 4     & 3     & 2     & 3     & 4     & 4     & 1 \\
  	\midrule
  	\textbf{Total} & \textbf{20} & 13    & 9     & 17    & 15    & 9     & 11    & 13    & 10    & 7 \\
  	\bottomrule
  \end{tabular}%
  Note: \toolNameBshort = \toolNameB, \toolNameAshort = \toolNameA, EF. = EnFuzz, Af.$^\beta$ = Autofz$^\beta$, A+. = AFL++, Hg. = HonggFuzz, Lib. = Libfuzzer, Ang. = Angora, and Rad. = Radamsa.% represents HonggFuzz, Libfuzzer, and Radamsa, respectively.
	\end{threeparttable}
 % \vspace{-0.2cm}
\end{table} 

\begin{table}[t]
	\centering
	\scriptsize
	\caption{RQ4. Comparison results of branch and path for the ablation study on fuzzer-test-suite for ten times}\label{tab:ns_google}
    \resizebox{\textwidth}{!}{
	\begin{threeparttable}
    
\begin{tabular}{l|lllll|lllll}
 \\ \hline
\multirow{2}{*}{\textbf{Project}} & \multicolumn{5}{c|}{\textbf{\# Branch}} & \multicolumn{5}{c}{\textbf{\# Path}} \\
   & \textbf{\toolNameAshort} & \textbf{\toolNameC}(P) & \textbf{\toolNameD}(P) & \textbf{\toolNameE}(P) & \textbf{\toolNameF}(P) & \textbf{\toolNameAshort} & \textbf{\toolNameC}(P) & \textbf{\toolNameD}(P)                           & \textbf{\toolNameE}(P) & \textbf{\toolNameF}(P) \\ \hline
\rowsh \bf BS     & \textbf{4,155}                          & 3,144 (<0.01)                           & 3,519 (<0.01)                           &  3,714 (<0.01)                                                      & 3,200 (<0.01)                                                      & \textbf{7,529}                          & 6,815 (<0.01)                           & 7,152 (<0.01)                           & 6,832 (<0.01)                            & 5,674 (<0.01)                            \\
\bf CA                           & \textbf{285}                            & \textbf{285 (0.98)}                               & \textbf{285 (0.98)}                    & \textbf{285 (0.98)}        & 236 (<0.01)                                                                                                          & \textbf{262}                            & \textbf{262 (0.98)}                               & \textbf{262 (0.98)}                               & \textbf{262 (0.98)}                                & 140 (<0.01)                              \\
\rowsh \bf GU     & \textbf{3,679}                          & 1,875 (<0.01)                           & 3,088 (<0.01)                           & 2,669 (<0.01)                                                      & 1,386 (<0.01)                                                      & \textbf{4,498}                          & 3,779 (<0.01)                           & 4,228 (<0.01)                           & 3,430 (<0.01)                            & 3,305 (<0.01)                            \\
\bf LC                           & \textbf{4,705}                          & 3,623 (<0.01)                           & 4,198 (<0.01)                           & 3,935 (<0.01)                                                      & 2,803 (<0.01)                                                      & \textbf{2,489}                          & 2,073 (<0.01)                           & 2,398 (<0.01)                           & 2,159 (<0.01)                            & 1,562 (<0.01)                            \\
\rowsh \bf LA     & \textbf{13,902}                         & 10,793 (<0.01)                          & 13,081 (<0.01)                          & 12,670 (<0.01)                                                     & 10,003 (<0.01)                                                     & \textbf{31,046}                         & 25,763 (<0.01)                          & 29,704 (<0.01)                          & 26,143 (<0.01)                           & 19,923 (<0.01)                           \\
\bf LS                           & \textbf{625}                            & 622 (<0.01)                             & 625 (<0.01)                             & 514 (<0.01)                                                        & 594 (<0.01)                                                        & \textbf{352}                            & 203 (<0.01)                             & 313 (<0.01)                             & 194 (<0.01)                              & 160 (<0.01)                              \\
\rowsh \bf LX     & \textbf{21,744}                         & 17,365 (<0.01)                          & 19,357 (<0.01)                          & 20,295 (<0.01)                                                     & 11,976 (<0.01)                                                     & \textbf{35,819}                         & 33,135 (<0.01)                          & 34,796 (<0.01)                          & 33,001 (<0.01)                           & 16,720 (<0.01)                           \\
\bf OS-0.1                       & \textbf{4,685}                          & 4,428 (<0.01)                           & 4,417 (<0.01)                           & 4,161 (<0.01)                                                      & 4,316 (<0.01)                                                      & \textbf{4,498}                          & 4,139 (<0.01)                           & 4,220 (<0.01)                           & 4,012 (<0.01)                            & 3,395 (<0.01)                            \\
\rowsh \bf OS-0.2 & \textbf{4,258}                          & 3,967 (<0.01)                           & 3,819 (<0.01)                           & 4,225 (<0.01)                                                      & 3,171 (<0.01)                                                      & \textbf{4,903}                          & 4,059 (<0.01)                           & 4,713 (<0.01)                           & 3,931 (<0.01)                            & 2,767 (<0.01)                            \\
\bf OS-1.0                       & \textbf{9,039}                          & 8,378 (<0.01)                           & 8,673 (<0.01)                           & 8,629 (<0.01)                                                      & 6,715 (<0.01)                                                      & \textbf{4,738}                          & 4,631 (<0.01)                           & 4,689 (<0.01)                           & 4,100 (<0.01)                            & 2,789 (<0.01)                            \\
\rowsh \bf PC     & \textbf{54,735}                         & 49,965 (<0.01)                          & 51,226 (<0.01)                          & 49,852 (<0.01)                                                     & 40,657 (<0.01)                                                     & \textbf{87,192}                         & 80,330 (<0.01)                          & 85,385 (<0.01)                          & 79,827 (<0.01)                           & 44,886 (<0.01)                           \\
\bf PJ                           & \textbf{915}                            & 812 (<0.01)                             & 873  (<0.01)                            & 717 (<0.01)                                                        & 600 (<0.01)                                                        & 697                                     & 688 (<0.01)                             & 701 (0.60)                                        & \textbf{707 (0.45)}                                & 380 (<0.01)                              \\
\rowsh \bf RE     & \textbf{18,751}                         & 13,877 (<0.01)                          & 17,436 (<0.01)                          & 16,989 (<0.01)                                                     & 14,598 (<0.01)                                                     & \textbf{17,165}                         & 15,906 (<0.01)                          & 16,928 (<0.01)                          & 16,312 (<0.01)                           & 14,373 (<0.01)                           \\
\bf WF                           & \textbf{3,779}                          & 2,613 (<0.01)                           & 3,376 (<0.01)                           & 2,959 (<0.01)                                                      & 1,951 (<0.01)                                                      & \textbf{1,126}                          & 1,022 (<0.01)                           & 1,105 (<0.01)                           & 1,022 (<0.01)                            & 393 (<0.01)                              \\
\rowsh \bf FT     & \textbf{60,233}                         & 57,264 (<0.01)                          & 57,425 (<0.01)                          & 56,080 (<0.01)                                                     & 48,796 (<0.01)                                                     & \textbf{30,191}                         & 29,876 (<0.01)                          & 29,793 (<0.01)                          & 27,408 (<0.01)                           & 26,712 (<0.01)                           \\
\bf HB                           & \textbf{40,293}                         & 34,478 (<0.01)                          & 38,794 (<0.01)                          & 36,774 (<0.01)                                                     & 31,072 (<0.01)                                                     & \textbf{18,335}                         & 14,773 (<0.01)                          & 17,154 (<0.01)                          & 14,582 (<0.01)                           & 16,365 (<0.01)                           \\
\rowsh \bf JS     & \textbf{7,128}                          & 6,819 (<0.01)                           & 7,028 (<0.01)                           & 6,487 (<0.01)                                                      & 6,644 (<0.01)                                                      & 1,247                                   & \textbf{1,298} (<0.01) & 1,135 (<0.01)                           & 1,226 (0.02)                                       & 1,105 (<0.01)                            \\
\bf LJ                           & \textbf{18,625}                         & 13,146 (<0.01)                          & 15,316 (<0.01)                          & 13,031 (<0.01)                                                     & 12,974 (<0.01)                                                     & \textbf{2,936}                          & 2,639 (<0.01)                           & 2,832 (<0.01)                           & 2,565 (<0.01)                            & 2,243 (<0.01)                            \\
\rowsh \bf LP     & \textbf{6,025}                          & 5,120 (<0.01)                           & 5,297 (<0.01)                           & 5,082 (<0.01)                                                      & 3,974 (<0.01)                                                      & \textbf{766}                            & 714 (<0.01)                             & 715 (<0.01)                             & 709 (<0.01)                              & 254 (<0.01)                              \\
\bf LL                           & \textbf{63,415}                         & 60,007 (<0.01)                          & 62,588 (<0.01)                          & 61,701 (<0.01)                                                     & 59,648 (<0.01)                                                     & \textbf{10,546}                         & 9,829 (<0.01)                           & 9,933 (<0.01)                           & 9,516 (<0.01)                            & 8,265 (<0.01)                            \\
\rowsh \bf OT     & \textbf{5,520}                          & 5,125 (<0.01)                           & 5,277 (<0.01)                           & 5,214 (<0.01)                                                      & 2,533 (<0.01)                                                      & 1,498                                   & 1,489 (<0.01)                           & \textbf{1,510 (<0.01)}  & 1,456 (<0.01)                            & 448 (<0.01)                              \\
\bf SL                           & \textbf{3,259}                          & 2,974 (<0.01)                           & 3,014 (<0.01)                           & 3,029 (<0.01)                                                      & 2,354 (<0.01)                                                      & \textbf{617}                            & 561 (<0.01)                             & 613 (<0.01)                             & 560 (<0.01)                              & 330 (<0.01)                              \\
\rowsh \bf VB     & \textbf{14,006}                         & 12,034 (<0.01)                          & 13,523 (<0.01)                          & 11,895 (<0.01)                                                     & 11,877 (<0.01)                                                     & \textbf{1,658}                          & 1,612 (<0.01)                           & 1,635 (<0.01)                           & 1,519 (<0.01)                            & 1,150 (<0.01)                            \\
\bf WP                           & \textbf{45,741}                         & 42,007 (<0.01)                          & 44,396 (<0.01)                          & 39,407 (<0.01)                                                     & 40,820 (<0.01)                                                     & \textbf{6,019}                          & 5,850 (<0.01)                           & 5,903 (<0.01)                           & 5,322 (<0.01)                            & 4,101 (<0.01)                              \\ \hline
\textbf{Total}                                  & \textbf{409,502}                        & 360,721                                                                      & 386,631                                                                      & 370, 315                                                                     &  322,898                                                                      & \textbf{276,127}                        & 251,446                                                                      & 267,817                                                                      & 246,795                                            &  177,440                                          \\
\textbf{\increase}               & -                                       & \textbf{13.52\%}                                                             & \textbf{5.92\%}                                                              & \textbf{10.58\%}                                                             & \textbf{26.82\%}                                                             & -                                       & \textbf{9.82\%}                                                              & \textbf{3.10\%}                                                              & \textbf{11.89\%}                                   & \textbf{55.62\%}                                   \\ \hline
\end{tabular}
	\end{threeparttable}
    }

\end{table} 

\begin{table}[t]
	\centering
	\scriptsize
	\caption{RQ4. Comparison results of unique crashes triggered for the ablation study on fuzzer-test-suite for ten times}
    \label{tab:ns_google2}
    \resizebox{0.5\textwidth}{!}{
	\begin{threeparttable}
\begin{tabular}{l|lllll}
\hline
\textbf{Project}  & \textbf{ \toolNameAshort} & \textbf{ \toolNameC} & \textbf{ \toolNameD} & \textbf{ \toolNameE} & \textbf{ \toolNameF} \\ \hline
\rowsh \bf BS     & \textbf{1}                              & \textbf{1}                         & \textbf{1}                         & \textbf{1}                         & \textbf{1}                         \\
 \bf CA                           & \textbf{3}                              & \textbf{3}                         & \textbf{3}                         & \textbf{3}                         & 1                                  \\
\rowsh \bf GU     & \textbf{1}                              & \textbf{1}                         & \textbf{1}                         & \textbf{1}                         & 0                                  \\
 \bf LC                           & \textbf{2}                              & \textbf{2}                         & \textbf{2}                         & \textbf{2}                         & \textbf{2}                         \\
\rowsh \bf LA     & \textbf{1}                              & \textbf{1}                         & \textbf{1}                         & \textbf{1}                         & 0                                  \\
 \bf LS                           & \textbf{2}                              & 1                                  & 1                                  & 1                                  & 1                                  \\
\rowsh \bf LX     & \textbf{3}                              & 2                                  & \textbf{3}                         & \textbf{3}                         & 1                                  \\
 \bf OS-0.1                       & \textbf{4}                              & \textbf{4}                         & 3                                  & 2                                  & 1                                  \\
\rowsh \bf OS-0.2 & \textbf{6}                              & \textbf{6}                         & \textbf{6}                         & \textbf{6}                         & 2                                  \\
 \bf OS-1.0                       & \textbf{6}                              & 4                                  & 5                                  & 4                                  & 3                                  \\
\rowsh \bf PC     & \textbf{7}                              & 6                                  & \textbf{7}                         & 6                                  & 2                                  \\
 \bf PJ                           & \textbf{3}                              & 2                                  & 2                                  & 1                                  & 0                                  \\
\rowsh \bf RE     & \textbf{1}                              & \textbf{1}                         & \textbf{1}                         & \textbf{1}                         & 0                                  \\
 \bf WF                           & \textbf{0}                              & \textbf{0}                         & \textbf{0}                         & \textbf{0}                         & \textbf{0}                         \\
\rowsh \bf FT     & \textbf{0}                              & \textbf{0}                         & \textbf{0}                         & \textbf{0}                         & \textbf{0}                         \\
 \bf HB                           & \textbf{1}                              & \textbf{1}                         & 0                                  & \textbf{1}                         & 0                                  \\
\rowsh \bf JS     & \textbf{3}                              & 2                                  & \textbf{3}                         & \textbf{3}                         & 1                                  \\
 \bf LJ                           & \textbf{0}                              & \textbf{0}                         & \textbf{0}                         & \textbf{0}                         & 0                                  \\
\rowsh \bf LP     & \textbf{1}                              & 0                                  & \textbf{1}                         & 0                                  & 0                                  \\
 \bf LL                           & \textbf{3}                              & 2                                  & \textbf{3}                         & 2                                  & 2                                  \\
\rowsh \bf OT     & \textbf{3}                              & \textbf{3}                         & 2                                  & 2                                  & 1                                  \\
 \bf SL                           & \textbf{3}                              & 2                                  & \textbf{3}                         & 2                                  & 1                                  \\
\rowsh \bf VB     & \textbf{4}                              & 3                                  & \textbf{4}                         & 3                                  & 2                                  \\
 \bf WP                           & \textbf{0}                              & \textbf{0}                         & \textbf{0}                         & \textbf{0}                         & \textbf{0}                         \\ \hline
\rowsh \textbf{Total}                                  & \textbf{58}                             & 47                                 & 52                                 & 45                                 & 21                                 \\
\textbf{ \increase}               & -                                       & \textbf{23.40\%}                   & \textbf{11.54\%}                   & \textbf{28.89\%}                   & \textbf{176.19\%}                  \\ \hline
\end{tabular}
	\end{threeparttable}
    }
\end{table}

Figure~\ref{fig:coverage_real} shows that \toolNameB achieves the best branch and path coverage, compared with its base fuzzers. 
Autofz$^\beta$ and AFL++ achieve similar branch coverage, while Autofz$^\beta$ achieves slightly better path coverage.
Note that AFL++ incorporates all the best features developed for the AFL-family fuzzers over the years, including AFLFast and FairFuzz~\cite{aflpp}.
Thus, AFL++ can be regarded as a fuzzer with \emph{internal} ensemble fuzzing strategies.
Nevertheless, \toolNameB can further improve the effectiveness of AFL++ by combining it with other base fuzzers, detecting 36.36\% more bugs with ~4\% and ~3\% improvement in branch and path coverage, respectively.

Figure~\ref{fig:fuzzer_time_real} further shows the average progressive resource scheduling results for base fuzzers of \toolNameB, which validate the effectiveness of our design.
Radamsa has the lowest coverage and fewest bugs during the entire ensemble fuzzing campaign (as shown in Table~\ref{tab:bug_real} and Figure~\ref{fig:coverage_real}). Therefore, its test resources decrease rapidly over time, eventually remaining within a low percentage range (0-5\%).
Conversely, \toolNameB quickly schedules more resources for AFL++ during the early stage.
As the campaign progresses into the later stage, the scheduler gradually shifts its focus to rotating base fuzzers other than Radamsa. Resources are allocated to hybrid fuzzers like QSYM and Angora to leverage their sophisticated constraint-solving capabilities for exploring deeper code regions. This confirms that the scheduling algorithm successfully balances early-stage efficiency with later stage exploration.

\greyboxb{Answer to RQ3:} {
\toolNameB demonstrates strong real-world effectiveness. Using state-of-the-art base fuzzers, it successfully detected 20 unique bugs in the newest versions of the nine projects, including 3 known CVEs and 5 previously unknown vulnerabilities, outperforming all evaluated baselines and base fuzzers.}

\subsection{RQ4-Ablation Study}

\noindent {\bf The impact of the Dynamic Resource Scheduling.}
Table~\ref{tab:ns_google} and Table~\ref{tab:ns_google2} show that Dynamic Resource Scheduling significantly contributes to the performance of \toolNameA.
Compared with \toolNameA, \toolNameC (Random Scheduling) covers 13.52\% fewer branches and triggers 23.40\% fewer unique crashes on average.
For \toolNameE  (Equal Allocation), which allocates resources evenly among each base fuzzer, the covered branches and unique crashes decrease by 10.58\% and 28.89\%, respectively.
This demonstrates that \toolNameA's MAB-based scheduler is significantly more effective than both random and static allocation strategies.
Moreover, \toolNameA is statistically differentiated (p-value $<$ 0.01) from \toolNameC on all projects except for c-ares, where all strategies quickly reach the same branch coverage (p-value = 0.98). 
Similarly, \toolNameA significantly outperforms \toolNameE (p-value $<$ 0.01) on almost all projects, with the exception of \textit{c-ares} (p-value = 0.98) and \textit{proj4} in terms of executed paths (p-value = 0.45).

\noindent {\bf The impact of the Multidimensional Seed Evaluation.}
The proposed multidimensional seed evaluation is another key performance driver.
Without it, \toolNameD (which only rewards seeds that cover unique execution paths, similar to Autofz) covers 5.92\% fewer branches and detects 11.54\% fewer unique crashes, as shown in Table~\ref{tab:ns_google} and Table~\ref{tab:ns_google2}.
\toolNameD is statistically differentiated from \toolNameA in terms of coverage on nearly all projects, confirming the stability of our multi-metric integration.

\noindent {\bf The impact of the Seed Synchronization.}
As shown in Table~\ref{tab:ns_google} and Table~\ref{tab:ns_google2}, \toolNameF (without seed synchronization) exhibits the most drastic performance decline, with branch coverage dropping by 26.82\% and the number of unique crashes falling from 58 to 21.
The poor performance of \toolNameF highlights the high cost of redundant exploration.
Without a shared global pool, independent fuzzer instances repeatedly waste CPU cycles on shallow program states already explored by others.
We think that seed synchronization is a prerequisite for ensemble fuzzing, which is used by EnFuzz and Autofz.
Note that the parallel mode of base fuzzers in RQ2 also uses seed synchronization.

\greyboxb{Answer to RQ4:} {
Ablating the dynamic resource scheduling (variants \toolNameC, \toolNameE), multidimensional seed evaluation (variant \toolNameD), or seed synchronization (variant \toolNameF) results in a significant decline in branch coverage, path execution, and unique crashes triggered.
The results demonstrate that while seed synchronization provides a collaborative foundation, the synergistic combination of our dynamic scheduler and multidimensional evaluation is essential for improving the effectiveness of ensemble fuzzing.}

\section{Discussions}
{\bf \toolName vs. Existing Ensemble Frameworks.}
\toolName differs conceptually and practically from prior ensemble frameworks in both its evaluation granularity and scheduling mechanism.
First, Autofz schedules resources by ``looking-forward'': regardless of how each base fuzzer performed previously, Autofz conducts a Preparation Phase at each round.
Meanwhile, \toolName schedules resources in a ``look-backward'' fashion: it evaluates the performance of fuzzers in previous rounds with a multidimensional evaluation strategy and accordingly schedules resources without profiling.
By looking-forward, Autofz can get precise performance estimations but requires running every fuzzer at the Preparation Phase of every round, including those performing poorly throughout the fuzzing campaign (as in evaluations~\cite{fu2023autofz}, more than 15\% of overall resources were spent on these fuzzers), leading to severe resource waste. 
In contrast, \toolName utilizes the valuable data from previous rounds to obtain an estimation with similar precision without conducting profiling, and more resources are scheduled for the better-performing fuzzers, which is the major reason for its improvement over Autofz.
Furthermore, the performance of the fuzzer can change significantly when new seeds are added to the corpus, which may be one reason why Autofz has the Preparation Phase.
However, due to the randomness of fuzzing, the improvements brought by new seeds are difficult to reflect in various metrics within a short preparation window.
Therefore, the Preparation Phase incurs high fixed costs without guaranteeing prediction accuracy.
To address this, \toolName combines the MAB mechanism with Softmax. First, the Softmax function ensures that fuzzers with lower scores still have a non-zero selection probability.
The exploration value is inversely proportional to the square root of the previously allocated resources (t).
If a fuzzer is ignored, its exploration value surges, dynamically preserving the possibility of selecting a poorly performing fuzzer to test whether a new seed can revitalize it.
It's important to note that, from a macro perspective, higher-performing fuzzers still receive more resources, as shown in Figure~\ref{fig:fuzzer_time_real}.
Secondly, to respond to the performance fluctuations brought about by new seeds and to avoid wasting resources when selecting a poorly performing fuzzer, \toolName introduces a lightweight active monitoring mechanism.
If a fuzzer fails to generate a valid seed within a preset time, the fine-tuning mechanism stops that fuzzer and reallocates resources to the currently higher-performing fuzzer.
Compared to EnFuzz, which treats ensemble fuzzing as a parallel execution problem with static resource allocation fixed at startup, \toolName formulates resource scheduling as a dynamic MAB problem.
It continuously shifts CPU cores to the more effective fuzzers tailored to the current program state.
Finally, while frameworks like CollabFuzz~\cite{collabfuzz} focus primarily on optimizing seed routing and synchronization logic among collaborative instances, \toolName uniquely addresses dynamic test resource scheduling based on continuous multidimensional fuzzer evaluation.

\noindent {\bf Seed Evaluation of \toolName vs. Power Schedules.}
Traditional fuzzers, such as AFLsmart~\cite{pham2019smart} and AFL++~\cite{aflpp}, utilize power schedules to evaluate the quality (e.g., whether passing programmatic parse) of an individual seed in their queues.
The primary goal of the power schedule is to allocate mutation energy, which determines how many times a fuzzer should mutate a specific seed based on metrics (e.g., execution time).
In contrast, the multidimensional seed evaluation of \toolName serves a framework-level objective tailored specifically for ensemble fuzzing. Instead of allocating mutation energy within a single fuzzer, \toolName evaluates seeds to find high-performance base fuzzers, scheduling resources dynamically. Furthermore, \toolName adjusts the weights of its metrics based on their standard deviations across all running base fuzzers.
This allows \toolName to emphasize metrics that best distinguish the performance gaps among diverse fuzzers, which is inapplicable to the localized power schedules of a single fuzzer. 
Additionally, to maintain consistency, these metrics are also used for seed synchronization, and seeds that contributed to these metrics are added to the global seed pool.

\noindent {\bf Robust Evaluation of \toolName.}
The result of fuzzing has randomness, raising potential concerns that dynamic schedulers might be misled by noise.
To evaluate \toolName's robustness under the influence of noise, we analyzed the variance across 10 independent runs.
Figure 3 shows the 80\% confidence intervals (i.e., the shaded areas) for all metrics over the 12-hour campaign. 
If the scheduler is sensitive to noise, the performance trajectory may exhibit high variance and erratic shifts across different runs.
Instead, the 12-hour time-series data demonstrates that \toolName has low variance, particularly in the number of covered branches and executed paths, where the relative error margins remain consistently minimal (i.e., around 1\% to 1.5\%).
This low variance empirically validates the robustness of \toolName.

\noindent {\bf Analysis of Metric Normalization.}
As described in Section~\ref{sec:sync}, \toolName aggregates multidimensional metrics using unnormalized incremental values.
To evaluate the impact of introducing normalization on scheduling efficiency, we conduct a comparative experiment on boringssl. 
Specifically, we construct a normalized variant of \toolName in which we apply standard Min-Max normalization to the per-round results ($c_0$ to $c_4$) before calculation.
We retain the dynamic standard-deviation-based weights to ensure they continue to represent the discriminative power of the metrics.
We then compare this variant against our default unnormalized approach across 10 independent runs. 
The results demonstrate that applying standard normalization does not yield a statistically significant improvement in fuzzing performance (Mann-Whitney U test p-values for both branch coverage and path coverage are $> 0.57$).

\noindent {\bf Threats to Validity.}
The representativeness of the selected test subjects can affect the fidelity of our conclusions.
To mitigate this threat, we selected 33 subjects in total, including all 24 of the widely adopted Google's fuzzer-test-suite and nine real-world projects.
These subjects are (1) large in size ($\sim$13,000 LoC on average), (2) well-maintained (containing thousands of revisions and issues), (3) popular (100+ stars on GitHub), and (4) diverse in functionalities (including libraries, utilities, and applications).
Furthermore, we observed that the fuzzer-test-suite is an older benchmark compared to the current standard, FuzzBench. 
We did not use FuzzBench because our primary baseline, EnFuzz, is deeply coupled with QSYM (a hybrid fuzzer currently unsupported by FuzzBench due to modern kernel compatibility issues). 
Note that the fuzzer-test-suite shares the majority of its real-world target subjects with FuzzBench (e.g., re2, libxml2, and openssl), ensuring that the subjects evaluated remain relevant to modern fuzzing challenges. 
To further mitigate the risk of evaluating on an older benchmark, we also added RQ3 to further evaluate \toolName on the latest upstream versions of widely used real-world projects.
Another threat is that the randomness of fuzzing and the execution of the subjects may be nondeterministic.
To mitigate this, we repeated all experiments ten times and conducted the Mann-Whitney U test~\cite{mcknight2010mann} to determine whether the results were statistically different.
\section{Related Work}\label{related}

{\bf Fuzzing Techniques.}
Fuzzing has been extensively studied, and various types of fuzzing techniques have been proposed, such as generation-based fuzzers, mutation-based fuzzers, and hybrid fuzzers.
Generation-based fuzzers utilize the input format specification and quickly generate massive test inputs that are syntactically correct.
The common types of input specifications include input models~\cite{spike,peach,skyfire,Guo_operand,smartgift} and context-free grammar~\cite{csmith,ifuzzer}, while some~\cite{radamsa,fuzzng} utilize domain-specific knowledge.
Mutation-based fuzzers~\cite{deep_rein,sfuzz,afl,aflfast,fairfuzz,libfuzzer,hfuzz,mopt,cmfuzz,cerebro,vuzzer,afl-hier,alpfuzz} generate new inputs by mutating existing ones guided by grey-box execution information, such as branch coverage and execution paths.
Finally, hybrid fuzzers run heavy code analysis tools such as concolic execution engines~\cite{driller, qsym, symcc} or taint analysis tools~\cite{intriguer, pata} alongside mutation-based fuzzers to improve overall fuzzing performance.
Recently, a popular trend has been to adopt machine learning to enhance fuzzing techniques.
Some researches focus on learning complex input formats~\cite{seqfuzzer,ganfuzz,degott2019learning,zakeri2021format},
while some fuzzers~\cite{deepfuzz,montage,titanfuzz,chatfuzz} train models to enhance the generation and mutation of input seeds.
Moreover, some fuzzers~\cite{fuzzguard,meuzz,omnifuzz,suzzer} use various machine-learning techniques to predict the effect of executing seeds and filter out those that are less likely to trigger interesting behaviors of the target program.
Finally, Angora~\cite{angora} uses taint analysis with a machine learning algorithm to generate inputs leading to
alternative paths.
All these fuzzers come with their strengths and weaknesses, and no fuzzer is in the dominant position~\cite{Zhu_roadmap_survey,CHEN_survey_2018118,Poncelet_so_many}.
\toolName is orthogonal to these works and can combine their strengths by including them as base fuzzers.

\noindent {\bf Parallel Fuzzing.}
Another research area similar to ensemble fuzzing is parallel fuzzing, which typically improves code coverage and vulnerability discovery by executing multiple independent or partially synchronized instances concurrently.
Xu et al.~\cite{xu2017designing} observe that AFL significantly slows down when it runs on 120 cores in parallel and design new operating primitives for fuzzing to improve the execution speed.
Liang et al.~\cite{liang2018pafl} propose a task partitioning method based on branch coverage bitmaps of the target program, with each fuzzer responsible for executing its assigned program branches.
Pham et al.~\cite{pham2021towards} assign seeds to different worker processes to drive fuzzing instances toward different regions of the program space.
Wang et al.~\cite{wang2021facilitating} introduce a task partitioning model based on edge coverage to ensure that parallel fuzzing instances execute mutually exclusive and balanced tasks.
Gu et al.~\cite{gu2022group} adopt a group-based corpus scheduling strategy, selecting high-quality seeds by combining energy and difference scores to improve testing efficiency and reduce task conflicts.
Chen et al.~\cite{chen2023mufuzz} restructure parallel fuzzing using a microservice architecture, alleviating synchronization blocking through concurrent microservices and reducing state synchronization via state partitioning.
Liang et al.~\cite{liang2024dodrio} optimize taint-analysis-based parallel fuzzing through real-time synchronization and load-balanced task dispatch, minimizing redundant behaviors and maximizing the utilization of computing resources.
This related research primarily focuses on task allocation and state synchronization to prevent redundant computation and resource waste.
Although existing parallel fuzzing tools can manage multiple fuzzer instances, they do not conduct in-depth research on test resource scheduling, which relies on testers to configure manually.
In contrast, \toolName uses test resources (e.g., CPU cores) as scheduling units. It dynamically allocates these test resources to different base fuzzers based on their performance. While Kraken's scheme~\cite{zhou2025kraken} for dynamically optimizing parallelism (to determine the number of running instances) also uses test resources as scheduling units, its core task is to analyze the optimal number of instances.
The task of \toolName is to analyze the percentage of test resources obtained by each base fuzzer.
Furthermore, regarding evaluation mechanisms, parallel fuzzing frameworks evaluate seeds (using a single metric such as branch coverage) to determine if the seed is sufficiently effective to be synchronized across instances or further modified.
They do not evaluate the relative effectiveness of individual fuzzing instances.
\toolName, on the other hand, evaluates the fuzzers themselves globally and comparatively by formulating the allocation as an MAB problem. By employing an unnormalized, multidimensional seed evaluation strategy (incorporating dimensions such as deep edges and unique crashes), Legion dynamically calculates the standard deviation of these metrics across all active instances. This dynamically adjusts the weight of each metric, ensuring resources are distributed based on true comparative performance rather than isolated, local seed value.

\noindent {\bf Vulnerability Detection.}
Software vulnerability detection is a crucial area for ensuring software security.
With software vulnerability detection, developers can implement remediation and mitigation measures before these vulnerabilities are exploited~\cite{harzevili2023survey}.
Zhang et al.~\cite{zhang2023vulnerability} proposed decomposing the control flow graphs of the code snippet into multiple execution traces, using a pre-trained code model and a convolutional neural network to learn path representations with intra- and inter-path attention, and inputting these into a classifier for vulnerability detection.
Wang et al.~\cite{wang2024combining} combine static source code information and dynamic program execution traces to represent code snippets and train a neural network model for vulnerability detection.
GRACE~\cite{lu2024grace} enhances LLM-based software vulnerability detection by combining graph structure information in code with contextual learning.
Additionally, Marashdih et al.~\cite{marashdih2023enhanced} propose a static taint analysis method that detects input validation vulnerabilities by analyzing source code and tracking tainted inputs.
Fuzzing is one of the mainstream methods for vulnerability detection.
Compared with these analysis methods, fuzzing has a lower false positive rate and can provide POCs to help developers quickly discover and fix vulnerabilities~\cite{pan2022automated, simsek2025pocgen, jiang2022evocatio}.
\toolName effectively combines multiple basic fuzzers to improve the efficiency of fuzzing.
	
\section{Conclusion}
\label{conclusion}

We propose a novel ensemble fuzzing framework named \toolName that dynamically schedules resources during an ensemble fuzzing campaign by adaptively learning base fuzzers' performance and evaluating the benefits of seeds from multiple dimensions.
The evaluation results show that
\toolName outperforms the state-of-the-art techniques EnFuzz~\cite{enfuzz} and Autofz~\cite{fu2023autofz} with the same set of base fuzzers, 
and detected 20 bugs in nine real-world projects with state-of-the-art base fuzzers, including three CVEs and five previously unknown bugs.

Our future work will focus on three directions.
The first one is prediction-based resource scheduling.
Due to the randomness of fuzzing, resource scheduling based on performance still cannot guarantee complete accuracy.
We plan to analyze the target program's code to make preliminary predictions about the performance of each base fuzzer and schedule resources accordingly.
The second direction is to reduce the cost of performance evaluation and resource scheduling.
Currently, at the end of each round, \toolName stops all base fuzzers and uses the seed generated by each base fuzzer to evaluate its performance.
Such a process costs extra execution resources.
One possible method to tackle this is to evaluate and update the performance of each base fuzzer on-the-fly during fuzzing, and only stop those fuzzers with no resources scheduled for the next round.    
The third direction is adapting the Multi-Armed Bandit model to the highly non-stationary nature of fuzzing. Specifically, we plan to introduce a time-decaying weight mechanism to the historical reward calculation. By assigning exponentially higher weights to the scores of the most recent rounds, the scheduler will be able to discount outdated performance history, allowing it to respond even more rapidly to sudden performance shifts triggered by newly synchronized seeds.

\section{Acknowledgments}
This research is supported by National Key R\&D Program of China (No. 2024YFB4506400) and sponsored by CCF-Huawei Populus Grove Fund.

\newpage
\bibliographystyle{ACM-Reference-Format}
\bibliography{bib2023}
\end{document}